\documentclass[traditabstract]{aa}
\usepackage{graphicx}
\usepackage{txfonts}
\usepackage{ulem}
\usepackage{natbib}
\usepackage{array}
\usepackage{xspace}

\newcommand{\kms}{km s$^{-1}$\xspace}

\begin{document}
\titlerunning{H$_2$ near-IR emission from Herbig Ae/Be stars.}
\title{A survey for near-infrared H$_2$ emission in Herbig Ae/Be stars: emission from the outer disks of HD 97048 and HD 100546
\thanks{Based on observations collected at the European Southern Observatory, Paranal, Chile. (Program IDs 079.C-0860C, 080.C-0738A, 081.C-0833A)}}

 \author{A. Carmona\inst{1,2}
\and G. van der Plas\inst{3,4}
\and M.E. van den Ancker\inst{4}
\and M. Audard\inst{1,2}
\and L.B.F.M Waters\inst{5,3}
\and D. Fedele\inst{6}
\and B.~Acke\inst{7}\fnmsep\thanks{Postdoctoral Fellow of the Fund for Scientific Research, Flanders.}
\and E. Pantin\inst{8}
}

\offprints{A. Carmona, \email{andres.carmona@unige.ch}}

\institute{ISDC Data Centre for Astrophysics, University of Geneva, chemin d'Ecogia 16, 1290 Versoix, Switzerland
 \and Observatoire de Gen\`eve,  University of Geneva, chemin des Maillettes 51, 1290 Sauverny, Switzerland
  \and Sterrenkundig Instituut 'Anton Pannekoek', University of Amsterdam, Kruislaan 403, 1098 SJ Amsterdam, The Netherlands
  \and European Southern Observatory, Karl-Schwarzschild-Str.2, D 85748 Garching bei M\"unchen, Germany
  \and SRON Netherlands Institute for Space Research, Sorbonnelaan 2, 3584 CA Utrecht, The Netherlands
  \and Department of Physics and Astronomy, Johns Hopkins University, Baltimore, MD 21218, USA
  \and Instituut voor Sterrenkunde, K. U. Leuven, Celestijnenlaan 200D, 3001 Leuven, Belgium
  \and Service d'Astrophysique, CEA Saclay, F-91191 Gif-sur-Yvette, Cedex, France
}

   \date{}

\abstract{
We report on a sensitive search for H$_2$ 1-0 S(1), 1-0 S(0) and 2-1 S(1) ro-vibrational
emission at 2.12, 2.22 and 2.25 $\mu$m in a sample of 15 Herbig Ae/Be stars
employing  CRIRES, the ESO-VLT near-infrared high-resolution spectrograph, at R$\sim$90,000.
We report the detection of the H$_2$ 1-0 S(1) line toward HD 100546 and HD 97048.
In the other 13 targets, the line is not detected. 
The H$_2$ 1-0 S(0) and 2-1 S(1) lines are undetected in all sources. 
These observations are the first detection of near-IR H$_2$ emission in HD 100546.
The H$_2$ 1-0 S(1) lines observed in HD~100546 and HD~97048 are observed at a velocity consistent with the rest velocity of both stars,
suggesting that they are produced in the circumstellar disk. 
In HD~97048, 
the emission is spatially resolved and it is observed to extend at least up to 200 AU from the star.
We report an increase of one order of magnitude in the H$_2$ 1-0 S(1) line flux  
with respect to previous measurements taken in 2003 for this star,
which suggests line variability.
In HD 100546 the emission is tentatively spatially resolved and may extend at least up to 50 AU from the star.
Modeling of the H$_2$ 1-0 S(1) line profiles and their spatial extent with flat keplerian disks 
shows that most of the emission is produced at a radius larger than 5 AU.
Upper limits to the H$_2$ 1-0 S(0)/ 1-0 S(1)  and H$_2$ 2-1 S(1)/1-0 S(1) line ratios in HD 97048
are consistent with H$_2$ gas at T$>$2000 K and suggest that the emission observed may be produced by X-ray excitation.
The upper limits for the line ratios for HD~100546 are inconclusive.
Because the H$_2$ emission is located at large radii, for both sources a thermal emission scenario (i.e., gas heated by collisions with dust) 
is implausible.
We argue that the observation of H$_2$ emission at large radii may be indicative of an extended disk atmosphere at radii $>$ 5 AU.
This may be explained by a hydrostatic disk in which gas and dust are thermally decoupled or by a disk wind caused by photoevaporation.
}

   \maketitle

\begin{table*}
\centering
\caption{Coordinates, spectral types, distances, radial velocities (RV), and SED group \tablefootmark{a}  of the program stars. } 
\begin{tabular}{l c c c r r c  }          
\hline\hline  
Star    &	 $\alpha$ (J2000.0)  & $\delta$ (J2000.0) &	Sp. Type & $d$ [pc]& 	RV [km s$^{-1}$] &Group \tablefootmark{a} 	  \\
\hline
\object{HD 58647} \tablefootmark{b}		&	07:25:56.10	&	--14:10:43.5	&	B9IVe \tablefootmark{d}	   &280\tablefootmark{d} &	 ...	 		        &	II\\
\object{HD 87643} \tablefootmark{c}	   &	10:04:30.29	&	--58:39:52.1	&	B2e \tablefootmark{e}			&	...  &  ...			                &	I\\
\object{HD 95881}		&	11:01:57.62	&	--71:30:48.3	&	A1/A2III/IVe \tablefootmark{f}	 &  118 \tablefootmark{f} & +36$\pm2$ \tablefootmark{g}   &	II\\
\object{HD 97048}		&	11:08:03.32	&	--77:39:17.5	&	A0Vpshe \tablefootmark{d}		&	180 \tablefootmark{d}  & +18$\pm3$ \tablefootmark{h} & I\\
\object{HD 100546	} &	11:33:25.44	&	--70:11:41.2	&	B9Vne \tablefootmark{d}			&	103 \tablefootmark{d}  & +16$\pm2$ \tablefootmark{h}	 & I\\
\object{HD 101412}	&	11:39:44.46	&	--60:10:27.7	&	B9.5Ve \tablefootmark{e}		&	160 \tablefootmark{g}  &~$-$3$\pm2$ \tablefootmark{g}       & I/II\\
\object{HD 135344B}	&	15:15:48.43	&	--37:09:16.0	&	F4Ve \tablefootmark{f}				&	140 \tablefootmark{f}  &~$+$2.5$\pm1.5$ \tablefootmark{i}	       & II, TD\\
\object{HD 141569}	&	15:49:57.75	&	--03:55:16.4	&	A0Ve	\tablefootmark{d}	&	99 \tablefootmark{d}  &~$-$2$\pm2$ \tablefootmark{g}					 & II, TD \\
\object{HD 144432	} &	16:06:57.96	&	--27:43:09.8	&	A9IVe \tablefootmark{f} &	145 \tablefootmark{f} 	 &~+2$\pm2$ \tablefootmark{g}		 & II\\
\object{HD 150193	} &	16:40:17.92	&	--23:53:45.2	&	A1Ve \tablefootmark{d}				& 150 \tablefootmark{d} &~$-$6$\pm2$ \tablefootmark{g}				 & II\\
\object{51 Oph}				&	17:31:24.95	&	--23:57:45.5	&	B9.5Ve \tablefootmark{d}			&	130 \tablefootmark{d}  & $-$11$\pm3$ \tablefootmark{j}   & II\\
\object{HD 169142} 	&	18:24:29.78	&	--29:46:49.4	&	A5Ve \tablefootmark{f}			&	145 \tablefootmark{f} &~$-$3$\pm2$ \tablefootmark{g}			 & I\\
\object{R CrA}					&	19:01:53.65	&	--36:57:07.6	&	A1-F7e \tablefootmark{f}		&	130  \tablefootmark{f}  &~~~0$\pm2$ \tablefootmark{g}				 & II\\
\object{HD 179218}	&	19:11:11.25	&	+15:47:15.6	&	A0IVe \tablefootmark{d}		&	240 \tablefootmark{d}  &~$-$9$\pm2$ \tablefootmark{g} 		    & I\\
\object{HD 190073}   & 20:03:02.51  &+05:44:16.7  & A2IVpe \tablefootmark{e} 		&	$>$290  \tablefootmark{g} &~+3$\pm2$ \tablefootmark{g}    &II\\
\hline    
\end{tabular}
\flushleft
\tablefoot{
\tablefoottext{a} SED classification by Meeus et al. (2001). Group I sources are with ``flared" disks and Group II sources are ``self-shadowed" disks.
TD means transition disk. These stars lack or have a weak near-infrared excess. This is interpreted as evidence of a hole or gap in the inner disk.
Because the star shows evidence of accretion the gap refers only to the lack of small dust particles in the inner disk.
\tablefoottext{b} Manoj et al. (2002) argue that HD~58647 may not be a Herbig Ae/Be star but a classical B[e] star.  
Baines et al. (2006) suggest that  HD~58647 is a binary star based on spectropolarimetry of the $H_{\alpha}$ line. 
\tablefoottext{c}  The status of HD~87643 as a Herbig Ae/Be star is controversial. 
Oudmaijer et al. (1998) argue that HD 87643 is an evolved B[e] star (see also Kraus 2009). 
Millour et al. (2009) based on AMBER-VLTI observations reported a close companion at 34 mas. 
References for the spectral type, distance, and heliocentric radial velocity: 
\tablefoottext{d} van den Ancker et al. (1998); 
\tablefoottext{e} SIMBAD;
\tablefoottext{f} Acke et al. (2004) and references there in;
\tablefoottext{g} Acke et al. (2005) and references there in;
\tablefoottext{h} this work, see discussion Sect. 2.2;
\tablefoottext{i} M\"uller et al. (2011);
\tablefoottext{j} Kharchenko et al. (2007).
}  
\label{star_table}
\end{table*}

\section{Introduction}
Circumstellar disks around pre-main sequence stars are the birth places of planets.
Hence, the characterization of their physical properties is of paramount importance. 
At the time when giant planets are in formation, protoplanetary disks are rich in gas. 
However, there are relatively few observational constraints of the gas content in the disk,
in particular for the inner disk ($R<10$ AU), the region where planets are expected to form. 
To study the gas content in the disk,  molecular and atomic line emission is used. 
Because the disk has a radial temperature gradient, different transitions of diverse
gas tracers probe different radii in the disk
(see reviews by Najita et al. 2007 and Carmona 2010).
These emission lines are in general produced in the surface layers of the disk where the dust is optically thin.

By far the main constituent of the gas in protoplanetary disks is molecular hydrogen (H$_2$).
However, given its physical nature (H$_2$ is an homonuclear molecule that lacks a dipole moment),
H$_2$ transitions are very weak.
In consequence,  in contrast to other gas tracers (most notably CO), 
H$_2$ emission is harder to detect.
With the advent of space borne observatories in the UV and infrared and ground based high-resolution infrared spectrographs,
it has been possible to start the search and study of H$_2$ emission from disks around young stars
from the  UV  to the mid-IR.

In the UV, H$_2$ electronic transitions trace in emission hot H$_2$ gas in the disk that is photoexcited (``pumped") by Ly$\alpha$ photons  
 (e.g. Valenti et al. 1993; Ardila et al. 2002; Herczeg et al. 2006) 
 or excited by electrons that are generated by X-rays (e.g., Ingleby et al. 2009, France et al. 2011). 
In absorption, UV transitions trace cold, warm, and hot H$_2$ in the line of sight (e.g., Martin-Za\"idi et al. 2005, 2008a), 
either from the disk itself (e.g., flared disk, close to edge-on disk, disk wind) 
or the star's circumstellar environment (e.g., envelopes).
In the mid-IR,  H$_2$ pure-rotational transitions trace thermal emission of warm gas at  few hundred K
(e.g. Thi et al. 2001; Bitner et al. 2007; Martin-Za\"idi et al. 2007, 2008b, 2010; Lahuis et al. 2007; Carmona et al. 2008b).
In the specific case of the near-infrared, 
ro-vibrational transitions of H$_2$ trace thermal emission of hot H$_2$ 
at a  thousand K, or H$_2$ gas excited by UV or X-rays (e.g. Weintraub et al. 2000; Bary et al. 2003, 2008; Itoh et al. 2003; Ramsay Howat \& Greaves 2007; Carmona et al. 2007, 2008a).
Because gas at these temperatures is expected to be located at radii up to a few AU,
H$_2$ near-IR emission has the potential of tracing the gas in the terrestrial planet region of disks
if the observed emission is thermal.

H$_2$ near-IR emission from disks has been mainly studied toward the low-mass T Tauri stars,
where it has been detected preferentially in objects exhibiting signatures of a high accretion rate 
(i.e., large H$\alpha$ equivalent widths and $U-V$ excess) and in a few weak-line T Tauri stars 
with bright X-ray emission (see the discussion sections of  Carmona et al. 2007, 2008a, and Bary et al. 2008).
However, for intermediate mass young-stars, i.e., Herbig Ae/Be stars, 
there is to date only one reported detection of near-infrared H$_2$ emission at the velocity of the star,
namely in the star HD 97048 (Bary et al. 2008).  
Because Herbig Ae/Be stars are bright in the near-IR,
the detection of the faint H$_2$ lines is challenging, in particular at low spectral resolution.
In this paper, 
we present the results of a considerable effort that we undertook to search for H$_2$ 1-0 S(1), 1-0 S(0) and 2-1 S(1) emission
toward Herbig Ae/Be stars employing  CRIRES, the ESO-VLT near-infrared high-resolution spectrograph, 
at resolution $R\sim$ 90,000.\\

The paper is organized as follows. 
In Section 2 we describe our CRIRES observations and the data reduction techniques.
In Section 3 we present the resulting spectra and constrain the
excitation mechanism of the detected lines based on H$_2$ line ratios. 
We then analyze the observed H$_2$ lines in the context of a flat disk model.
In Section 4 we discuss our results.
First, we compare the H$_2$ line profiles detected with the line profiles of other disk gas tracers.
We then discuss our H$_2$ observations and observations of dust in the inner disk.
Thereafter, 
we analyze H$_2$ near-IR emission in the frame of diverse disk structure models.
We propose explanations for the relatively large number of non-detections,
and explore diverse scenarios to explain the detections of near-IR H$_2$ lines in HD 100546 and HD 97048.
Finally, in Section 5 we present our conclusions.
 
\section{Observations and data reduction}
\subsection{Observations}
We observed a sample of 15 nearby Herbig Ae/Be stars of diverse spectral types 
and disk geometries  based on the spectral energy distribution (SED)
classification introduced by Meeus et al. (2001) and SED modeling by Dullemond et al. (2001) 
(i.e., group I: flared disks, group II: self-shadowed disks).
We summarize the properties of the stars of our program in Table~\ref{star_table}. 

The sources were observed with the CRyogenic high-resolution 
InfraRed Echelle Spectrograph (CRIRES) mounted at the ESO-VLT UT1 (Antu) atop Cerro Paranal, 
Chile (K\"aufl et al. 2004). 
The observations are standard long-slit spectroscopy observations.
We employed a slit of width 0.2", with the slit rotated along the parallactic angle
resulting in a spectral resolution $R$ of 90,000 (or 3.2 km~s$^{-1}$).
We used adaptive optics (MACAO - Multi-Applications Curvature Adaptive Optics) 
to optimize the signal-to-noise ratio and the spatial resolution. 
The typical spatial resolution achieved was 0.22" (i.e., FWHM in the continuum).
To correct for the sky emission, 
we used  the standard nodding technique along the slit employing a nod throw of 10". 
To correct for telluric absorption and flux-calibrate the spectrum,
we observed spectrophotometric standard stars 
 immediately before or after the science observations with the same instrument 
 setup. 
Our observations were carried out in service and visitor mode in 2008 
under the ESO programs 080.C-0738A and 081.C-0833A. 
We complemented our data for HD~97048 with CRIRES data taken by one of us with a similar setup 
(ESO program 079.C-0860C, PI E. Pantin).
In these observations the slit was oriented in the north-south direction and we used a standard star observed 
in our program 080.C-0738A to perform the telluric correction and flux calibration.
In Table~\ref{table_observations} (in the appendix) 
we present a summary of the observations and the point-spread function (PSF) FWHM achieved in the target and the calibrator.

\subsection{Data reduction and calibration}
 The data were reduced using the CRIRES pipeline V1.7.0\footnote{http://www.eso.org/sci/data-processing/software/pipelines/index.html} up to extraction of the 1D spectrum,
then custom IDL routines were employed for the telluric correction, accurate wavelength calibration, barycentric correction, cosmic ray cleaning and flux calibration.
For each chip flat-fielded image pairs in the nodding sequence (AB) were subtracted and averaged producing a combined image frame, thereby
performing the sky-background correction.
Raw frames at each nodding position were corrected for random jittering, employing the information in the fits headers. 
The ensemble of combined frames were stacked in one single 2D image spectrum,
from which a one-dimensional spectrum was extracted by summing the pixels in the spatial direction inside the PSF after background subtraction.
The CRIRES pipeline provides a first wavelength calibration by cross-correlation with the Th-Ar lamp frame taken during the same night of the observations,
and the final product of the CRIRES pipeline is a 1D spectrum. 

To correct for telluric absorption, the 1D extracted science spectrum was divided by the 1D extracted spectrum of the standard star. 
The standard star spectrum was first corrected for differences in air-mass and air-pressure with respect to the science target spectrum employing 
\begin{center}
$I_{STD_{corrected}} =I_{0}\; \exp{ \left(
-\tau \;
\frac{\displaystyle \bar{X}_{TARGET}}{\displaystyle \bar{X}_{STD}} \; 
\frac{\displaystyle \bar{P}_{TARGET}}{\displaystyle \bar{P}_{STD}} \right)~_.}
$
\end{center}
Here $\bar{X}$ is the average of the airmass and 
$\bar{P}$ is the average of the air pressure.
The continuum $I_{0}$ has been defined as uniform with a value equal to the mean plus 
three standard deviations
($I_{0}= \bar{I}_{\rm obs} + 3\sigma_{I_{\rm obs}}$). 
The optical depth $\tau$ is derived from
the measured data using $\tau = -\ln{\left(I_{obs}/{I_{0}}\right)}$.
There is some freedom in the choice of the continuum level,
we selected the mean plus 3$\sigma$ to set the continuum over the noise level,
which resulted in $I_{obs}/{I_{0}}<1$ along the spectrum.
The result of this correction is to make the depth of the telluric lines in the standard star spectrum
similar to the depth of the telluric lines in the science spectrum.
Small offsets of a fraction of a pixel in the wavelength direction were applied to the standard star spectrum 
until the best telluric correction (i.e., signal-to-noise) in the corrected science spectrum was obtained.

By comparing the sky absorption lines in the non-telluric corrected 1D science spectrum produced by the CRIRES pipe-line and 
a HITRAN model of Paranal's atmosphere, a final wavelength calibration was established. 
The typical wavelength accuracy achieved is 0.5 km s$^{-1}$. 
The spectra were corrected for the radial velocity (RV) of the star and the motion of the Earth-Moon-Sun system at the moment of the observation
using the heliocentric velocity correction given by the IRAF\footnote{http://iraf.noao.edu/} 
task {\it rvcorrect} and heliocentric radial velocities from the literature (see Table \ref{star_table}).

For HD~100546 and HD~97048, two measurements of the radial velocity are available from the literature.
Donati et al. (1997) estimate an RV of 17$\pm$5 \kms for HD~100546.
Acke et al. (2005) derived an RV of 18 \kms for  HD~100546, and of 21 \kms for HD~97048.
To obtain an independent estimate of the radial velocity of the objects, we used two additional methods.
First, we used archival high-resolution FEROS\footnote{Fiber-fed FEROS is the Extended Range Optical Spectrograph mounted at the ESO -- Max Planck 2.2m telescope at la Silla, Chile. It covers the complete optical spectral region in one exposure.}
(R$\sim$40000) spectra of HD~100546 and HD~97048.
We determined the radial velocity by fitting Gaussians to narrow photospheric absorption lines not affected by blending (because {\it v}\,sin\,$i$) or emission,
and comparing their centers to the center of Gaussians fitted to the same spectral features in rotationally broadened BLUERED (Bertone et al. 2008) 
high-resolution (R$\sim$500,000) synthetic spectral models corresponding to the spectral types of  HD~97048 and HD~100546.
We obtained for HD~100546 an RV of 16$\pm$2 \kms, and for HD~97048 an RV of 17$\pm$2.5 \kms.
Second, we derived the radial velocity from the center of the 
CO ro-vibrational lines observed at 4.7 $\mu$m in CRIRES R$\sim$ 90000 spectra taken by our team (van der Plas et al. 2009).
For this set of data we found an RV of  14$\pm$2 \kms for HD~100546 and an RV of  16$\pm$2 \kms for HD~97048.

Our results are consistent with the values of Donati et al. (1997)  and  the values of 
Acke et al. (2005), assuming a typical error of 2 \kms for the latter.
Donati et al. (1997)  derived their RV estimate for HD~100546 based only on the \ion{Mg}{ii} doublet at 4481 \AA.  
Because we used 12 additional photospheric lines in our determination of the  RV for HD~100546,
we consider our estimation to be more precise  (indeed, we obtained an RV of 17.3 \kms for the \ion{Mg}{ii} lines). 
The principal limitation on the derivation of the radial velocities using optical spectra is that we are dealing with 
B9 and A0 stars, and very few symmetric absorption lines are present in the spectra for the RV determination.
Hydrogen lines are affected by emission, and several lines are affected by blending owing to rotation.
For this reason, the independent estimation of RV from the CO emission lines is also important.

To apply the radial-velocity correction, we used the average value between of the three radial velocity determinations.
We employed an RV of 16$\pm$2 \kms for HD~100546, and an RV of 18$\pm$3 \kms for HD~97048.
 
Absolute flux calibration was made by multiplying the telluric corrected spectrum by the flux of a Kurucz model of the spectral type 
of the standard star at the observed wavelengths. 
The absolute flux calibration is accurate at the 20--30\% level. 
Imperfections in the telluric correction and, most importantly, slit losses owing to the narrow slit and AO (Adaptive Optics) performance are the principal sources of uncertainty.
The data of the settings $\lambda_{ref}$=2117 nm and $\lambda_{ref}$=2123 nm in the 2008 observations
were taken with the same exposure time (see Table \ref{table_observations}). 
Their combined spectrum was derived calculating the weighted average using  the continuum flux as weight.
The 2007 and 2008 observations of HD~97048 have different exposure times  (1920s and 320s respectively).
The final combined spectrum of HD~97048 is the weighted average using the exposure time as weight.
 
\subsection{Position-velocity diagrams}
To construct position-velocity diagrams and to compare the 2D spectrum with 2D disk models,
the 2D spectrum of the target  was further processed: 
{\it (i)} it was corrected for the trace in the dispersion direction by fitting a second-order polynomial 
to the PSF centers (obtained by a Gaussian fit) as a function of the wavelength, and then shifting each column of pixels such that 
the final trace is a straight line (correction on the order of 0.005 to 0.01 pixel); 
{\it (ii)} to correct for the telluric absorption, each row in the spatial direction was divided by the
1D spectrum of the standard star; 
{\it (iii)} the 2D spectrum was re-centered in the wavelength direction such that the center of the line profile is at 0 km s$^{-1}$ 
(in Sect. 3 we show that the center of the 1D line profile is at a velocity consistent with the rest velocity of the star. 
The re-centering correction of the 2D spectrum is smaller than the uncertainty of
the radial velocity, which justifies this step);
{\it (iv)} using the wavelength solution of the final extracted 1D science spectrum, the 2D spectrum was resampled in the
dispersion direction such that wavelength scale had a uniform sampling;
{\it (v)} the flux on each pixel was scaled in such a way that the extracted 1D spectrum has the continuum equal to 1; 
finally, {\it (vi)} using the distance of the star and the CRIRES pixel scale of 0.086 arcsec/pixel,
we derived the pixel scale in AU using as zero reference the pixel corresponding to the center of the PSF trace. 
In this way, a normalized, telluric corrected, wavelength and spatial calibrated 2D spectrum was obtained.

We have two observations  in two epochs with slits at different position angles for HD~97048.
Because the line observed in 2007 has a much better S/N than the line observed in 2008 (see Fig. \ref{HD97048_average}), 
we employed only the data taken in 2007 to model the line profile
and the analysis of the 2D spectrum (i.e., determination of the spatial peak position (SPP),
the FWHM of the PSF, and construction of the position-velocity diagrams).
The 2007 observations were taken with the slit in the N/S direction (see Table~\ref{table_observations} in the appendix).

Two spectra were taken immediately after each other for HD~100546  :
one with the wavelength setting centered at 2117 nm and another centered at  2123 nm.
We produced an averaged 2D spectrum
by making a cut at -50 to +50 km s$^{-1}$ of the H$_2$ 1-0 S(1) line in each observation
and averaging the 2D frames using their continuum flux as weight.
\begin{figure*}
 \centering              
\begin{tabular}{cc}
\includegraphics[height=0.2\textheight]{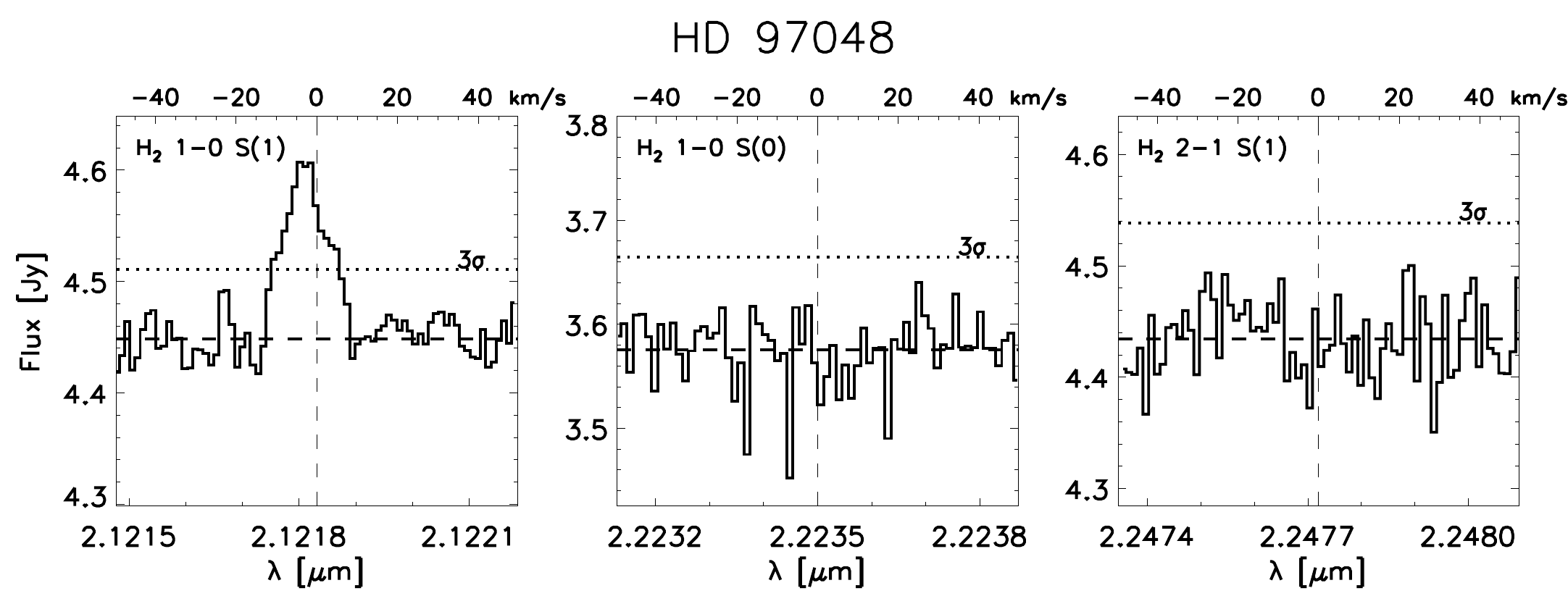} \\
\includegraphics[height=0.2\textheight]{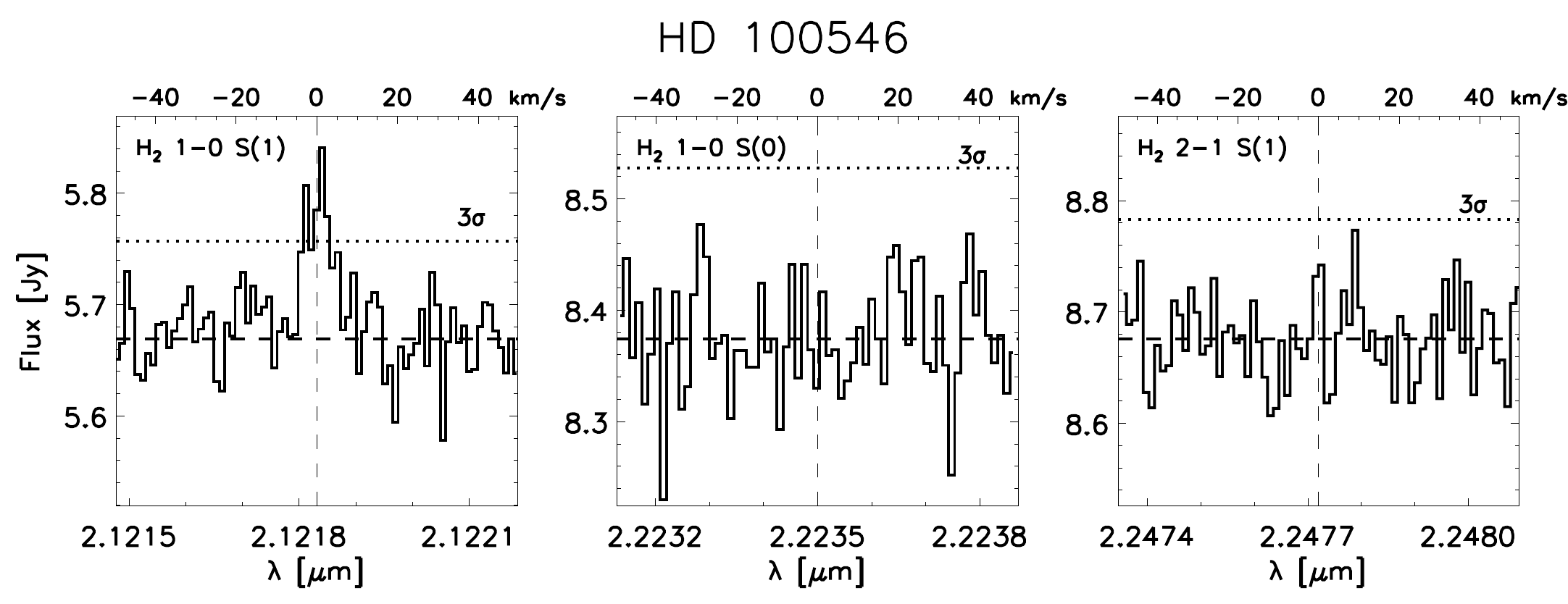} \\
\end{tabular}
 \caption{
Observed CRIRES spectrum at the location of the H$_2$ 1-0 S(1), 1-0 S(0) and 2-1 S(1) lines in HD~97048 and HD~100546, 
the only two sources displaying H$_2$ emission. 
Note that the H$_2$ 1-0 S(1) line in HD~97048 is the weighted average of the 2007 and 2008 observations (see Fig.~\ref{HD97048_average}). 
The spectra are presented at the rest velocity of the stars.
}
\label{fig_h2_hd100546_hd97048}
 \end{figure*}   
 
 \begin{figure*}
   \centering                  
\includegraphics[height=0.2\textheight]{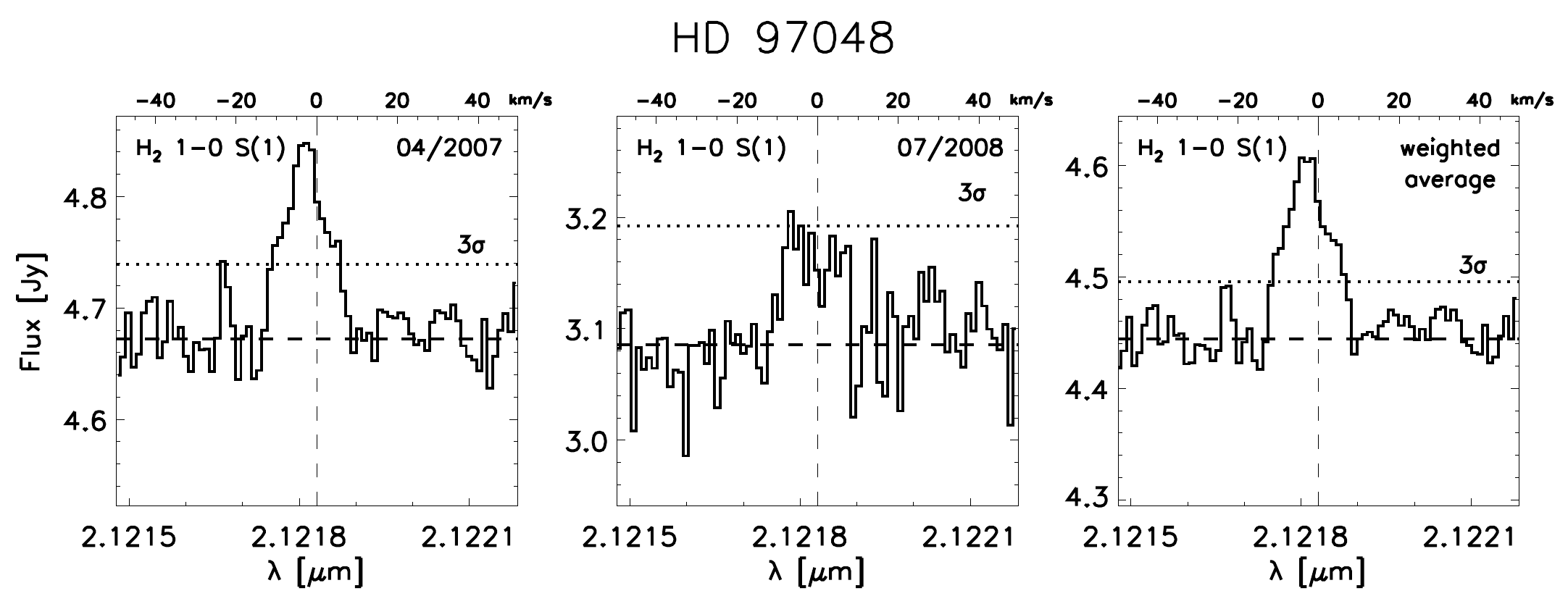}
\caption{HD~97048 H$_2$ 1-0 S(1) spectra observed in April 2007 (left), in July 2008 (middle), 
and (right) the weighted average spectrum using the exposure time as weight (1920s and 320s for the 2007 and 2008 observations respectively). 
All spectra are presented at the rest velocity of HD~97048.
}
\label{HD97048_average}
   \end{figure*}
   \begin{figure*}
\centering
\includegraphics[width=0.49\textwidth,angle=90]{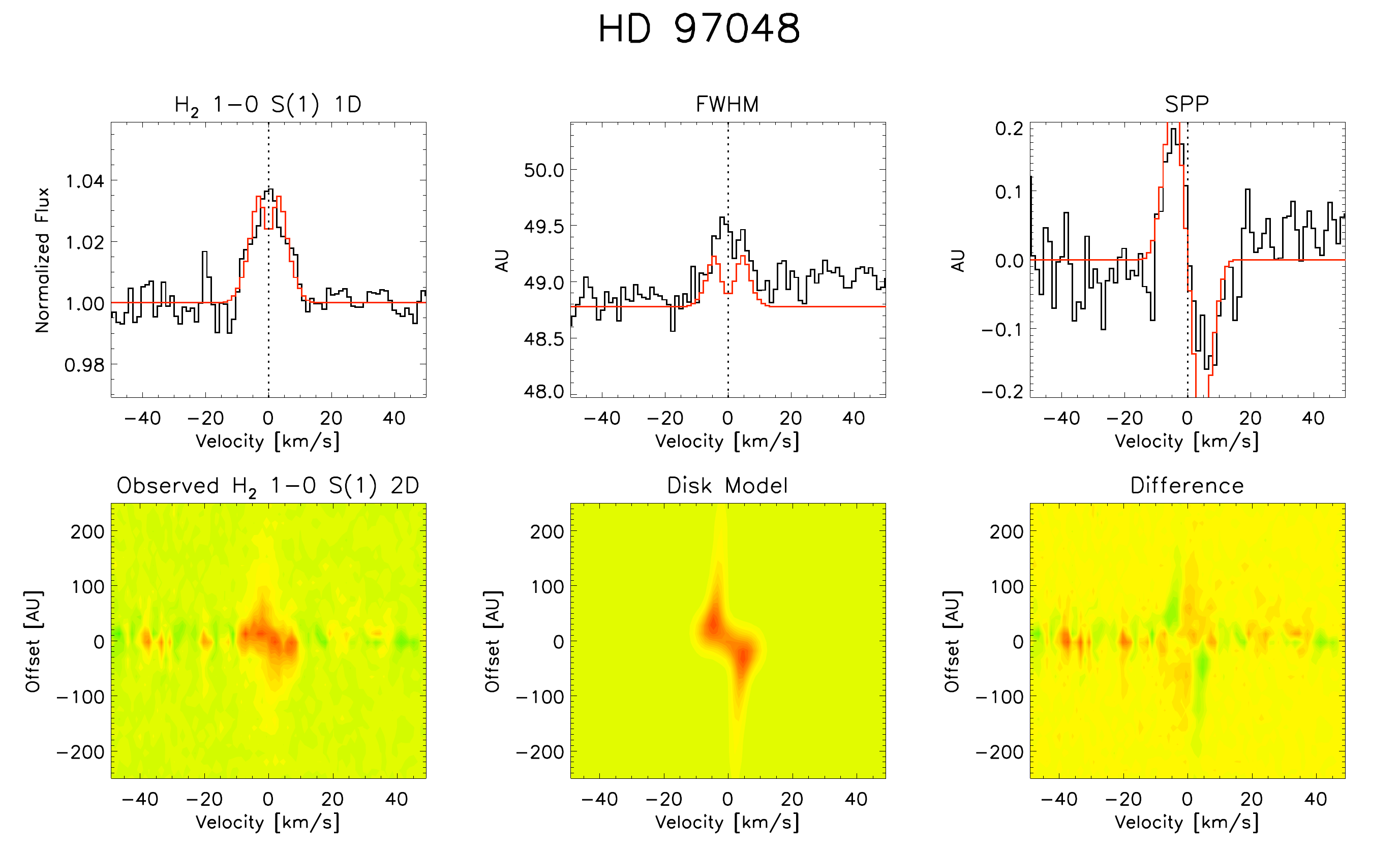} \\[2ex]
\includegraphics[width=0.49\textwidth,angle=90]{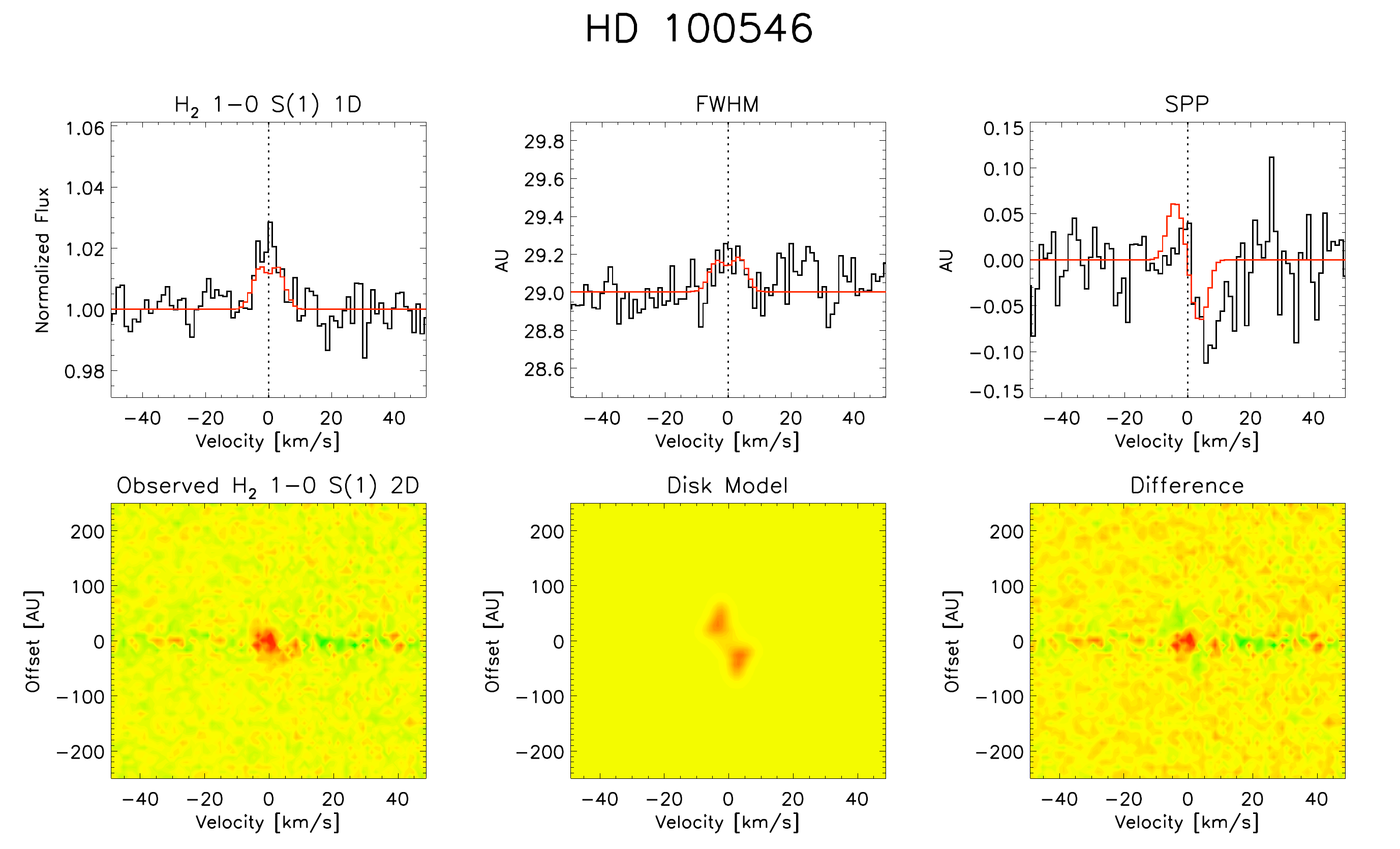} 
\caption{Observed and modeled spectrum for HD~97048 and HD~100546. 
For each star the {\it upper three panels} display the observed (in black) and the modeled (in red) H$_2$ 1-0 S(1) extracted 
1D line profile, the FWHM of the PSF, and the spatial peak position (SPP., i.e. center of Gaussian fit to the PSF).
The {\it lower three panels} show the position velocity diagrams of the 2D spectrum after subtraction of the continuum PSF. 
The first panel displays the observed 2D spectrum, 
the second shows the modeled 2D spectrum, 
and the third their difference.
We show a disk model with $\alpha=$1.5, $R_ {\rm in}$=9 AU, and $R_ {\rm out}$=400 AU for HD~97048 
and a disk model with $\alpha=$1.0 $R_ {\rm in}$=20 AU, and $R_ {\rm out}$=80 AU for HD~100546.
The spatial scale on the figures is 1 AU = 5.6 mas for HD~97048, and  1 AU = 9.7 mas for HD 100546.
Note that the H$_2$ lines have been shifted such that their center is at 0 km s$^{-1}$.
The 2007 and 2008 observations of HD~97048 have different slit PA.
Because the 2007 data have much better S/N than the 2008 data, 
we considered for this figure and line modeling only the 2007 data.
}
   \label{model_lines}
\end{figure*}  

\section{Results and analysis}
In Figs. \ref{fig_h2_hd100546_hd97048} and \ref{fig_nondetections} (in the appendix)
we present the results of our survey.
The H$_2$ 1-0 S(1) line is detected at the 5$\sigma$ level 
in HD~100546 and at the 8$\sigma$ level in HD~97048. 
In the other 13 targets the line is not detected. 
The H$_2$ 1-0 S(0) and 2-1 S(1) lines are undetected in all sources. 
We present in Table~\ref{upper_limits} the measured H$_2$ line fluxes and upper limits.

The  H$_2$ 1-0 S(1) lines observed in HD~97048 and HD~100546 are single-peaked.
For HD~97048,
a Gaussian fit to the line profile sets its center at $-2.9{\pm3}$ km s$^{-1}$ and the FWHM to $12.5\pm1$ km s$^{-1}$. 
Because the line FWHM is broader than the spectral resolution of CRIRES ($\sim3$ km s$^{-1}$),
the line is spectrally resolved.
In the 2007 2D spectrum, 
we observe that the FWHM of the observed PSF increases a few milli-arc-seconds at the position of the line, and
we detect a clear motion of the spatial peak position (SPP., i.e., center of the gaussian fit to the observed PSF) 
along the line, which indicates that the emission is spatially resolved.
The SPP displays the typical behavior for disk emission ruling out foreground emission (see Fig.~\ref{model_lines}).

For  HD~100546, 
the Gaussian fit to the H$_2$ 1-0 S(1) line profile sets the center 
at $0.5$${\pm2}$ km s$^{-1}$ and its FWHM to $8\pm1$ km s$^{-1}$.
The line is spectrally resolved.
In this source,
the FWHM of the observed PSF slightly increases at the position of the line,
which tentatively  indicates that the emission is spatially resolved (see Fig.~\ref{model_lines}).
The SPP signal displays a shape suggesting an extended emission component at positive velocities.
However, because the SPP signal is inside the 3$\sigma$ noise level,
it is difficult to establish at the S/N of our data whether this signal is real or noise residuals. 
The SPP signal observed is, nevertheless, consistent with the expected SPP signal of the rotating disk observed in CO (van der Plas et al. 2009)
at the position angle of our observations (PA$\sim27^\circ$). 
Additional observations with higher S/N are required to test this.

In Table \ref{line_properties} we summarize the parameters of the 1--0 S(1) H$_2$ lines detected in HD~97048 and HD~100546.
Taking into account an uncertainty of 2--3 km s$^{-1}$ in the radial velocity of HD~100546 and HD~97048,
the emission is observed at a velocity consistent with the rest velocity of the stars.
Because both sources have circumstellar disks, the most likely scenario is that the emission arises from the disk.
The data could also be consistent with emission of a disk-wind or outflow, but, in that case, it must have a 
low projected velocity ($<$ 6 km s$^{-1}$).
 
  \begin{table}
\setlength{\tabcolsep}{1.1mm}
\caption{Summary of the measured line fluxes and 3$\sigma$ upper limits.}              
\centering                                      
\begin{tabular}{l c c c c c c}          
\hline\hline                        
			&  	H$_2$ 1-0 S(1)                         & H$_2$ 1-0 S(0)	& H$_2$ 2-1 S(1)	\\
Star	   &  $\lambda=$2121.83 nm & $\lambda=$2223.50 nm &  $\lambda=$ 2247.72 nm\\	
\hline
HD~58647 & $< 2.5$  & $< 2.3$  & $< 2.3$ \\
HD~87643 & $< 12$  & $< 7.6$  & $< 9.4$ \\
HD~95881   & $< 3.5$  & $< 1.3$  & $< 1.1$ \\
{\bf HD~97048}   &  {\bf 9.6$\pm$2.9} &  $< 1.3$	  & $< 1.5$  \\
{\bf HD~100546} &	{\bf 5.4$\pm$1.6} & 	$< 2.1$	& $< 1.5$ \\
HD~101412 & $< 4.6$  & $< 1.6$  & $< 1.7$ \\
HD~135344B & $< 1.6$  & $< 1.7$  & $< 3.5$ \\
HD~141569  & $< 2.3$  & $< 0.5$  & $< 0.4$ \\
HD~144432 & $< 0.2$  & $< 0.2$  & $< 0.3$ \\
HD~150193 & $< 0.9$  & $< 3.0$  & $< 1.8$ \\
51 Oph			 & $< 2.7$  & $< 2.9$  & $< 2.7$ \\
HD~169142  & $< 0.9$  & $< 0.5$  & $< 0.7$ \\
R CrA				& $< 7.2$  & $< 9.4$  & $< 6.6 $ \\
HD~179218  & $< 0.9$  & $< 1.0$  & $< 0.9 $ \\
HD~190073 & $< 0.8$  & $< 0.4$  & $< 3.3$ \\
\hline
\end{tabular}
\flushleft
\tablefoot{ All line fluxes and upper limits given in $10^{-14}$ erg s$^{-1}$ cm$^{-2}$.
The uncertainty on the integrated line fluxes corresponds to an uncertainty of 30\% on the continuum flux.}
\\[0.5cm]
\label{upper_limits}
\end{table} 
\begin{table}
\setlength{\tabcolsep}{1.1mm}
\caption{Parameters of the 1-0 S(1) 	H$_2$ lines detected.}              
\centering                                      
{
\begin{tabular}{ l l c c c c c c}          
\hline\hline                        
  									& $\lambda_0$ & $\lambda_{\rm center}$\tablefootmark{a}  & $\delta_{\lambda}$\tablefootmark{b}  				& FWHM \\
Star & [nm]				&	 [nm]										& [km~s$^{-1}$]				& [km~s$^{-1}$]	\\
											\hline
HD~97048   &  2121.831		   & 2121.811$\pm0.028$   & -2.9$\pm3$  & 12.5$\pm1$ \\
HD~100546 &  2121.831 		& 2121.835$\pm0.014$   & 0.5$\pm2$  & 8$\pm1$ \\
\hline
\end{tabular}
}
\flushleft
\tablefoot{ 
\tablefoottext{a}\tablefoottext{b} The error in the center and $\delta_{\lambda}$ is the combined error in the wavelength calibration of 0.5 km s$^{-1}$ and 
the uncertainty in the radial velocity of the sources.}
\label{line_properties}
\end{table}
  \begin{figure}
 \centering              
\includegraphics[width=0.39\textwidth]{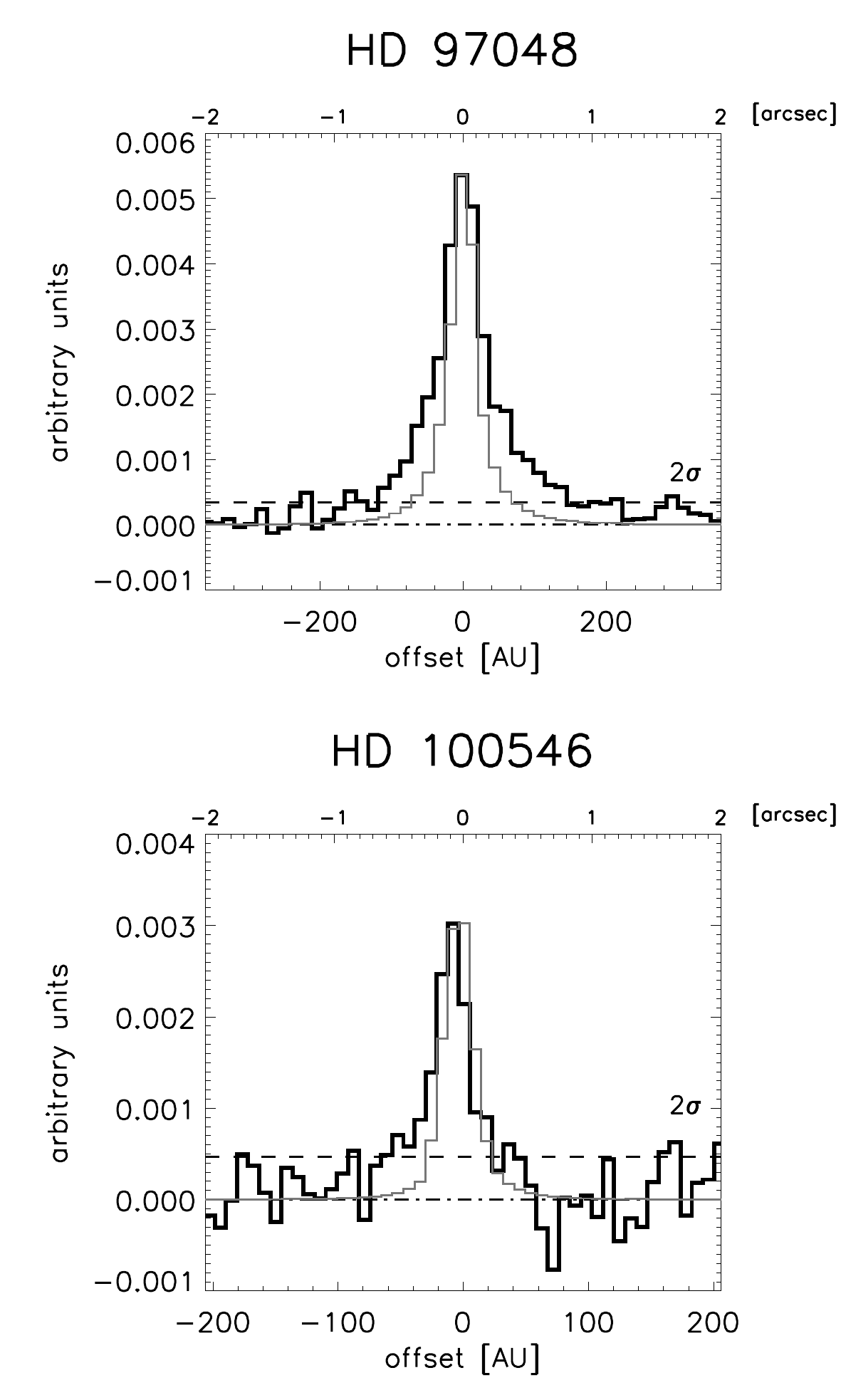} 
 \caption{
Sum of the counts from -15 to +15 km s$^{-1}$ after subtraction of the continuum PSF in the 2D spectrum.  
 The dashed line represents the 2$\sigma$ level of the background noise.
 The light gray line shows the scaled PSF of the continuum.
 }
\label{cumulative_PSF_counts}
 \end{figure} 
To quantify the spatial extent of the emission lines detected in HD~97048 and HD~100546, 
we subtracted from the 2D spectrum the average PSF observed in the continuum from -50 to -20 km s$^{-1}$ and +20 to +50 km s$^{-1}$ with respect to the line's position,
and produced two diagrams.
The first diagram is the position-velocity diagram of the 2D continuum-subtracted data (see Fig.~\ref{model_lines}). 
In the second diagram, we summed all the counts from -15 to +15 km s$^{-1}$ and plotted the resulting cumulative counts as a function of
 the spatial offset (see Fig.~\ref{cumulative_PSF_counts}).

In the case of HD~97048 these two diagrams show us that the H$_2$ emission extends at least up to -150 AU (at negative velocities) and +200 AU (at positive velocities). 
The position-velocity diagram displays the typical butterfly shape of a disk in rotation:
the line emission at negative velocity tends to have a positive spatial offset, and the line emission at negative velocity tends to have a negative spatial offset.
This is consistent with emission produced in a rotating disk and confirm the results suggested by the SPP.

The cumulative counts diagram of HD~100546 (see Fig.~\ref{cumulative_PSF_counts}) 
shows that the H$_2$ 1-0 S(1) line extends at least up to -50 AU to +40 AU.
The emission is slightly stronger at negative spatial offsets.
The position-velocity diagram displays the emission concentrated at zero velocities and shows that the emission is much more compact (i.e., less extended) than in
the case of HD~97048. At the S/N of the HD~100546 data, the signature of a rotating disk is not apparent, although
weak extended emission is observed at positive velocities (see Fig.~\ref{model_lines}).

Note in Fig.~\ref{cumulative_PSF_counts}
that for both stars 
the FWHM of the PSF at the position of the line (after subtracting the PSF of the continuum) 
is larger than the FWHM of the PSF of the continuum (overplot in light gray in Fig.~\ref{cumulative_PSF_counts}). 
For HD~97048 the continuum-subtracted PSF FWHM
at the H$_2$ line position is $\sim$520 mas, while the FWHM of PSF of the continuum is $\sim$255 mas.
For HD~100546 the continuum-subtracted PSF FWHM  is $\sim$400 mas, while the FWHM of the continuum PSF is $\sim$250 mas.
This justifies the approach of using Fig.~\ref{cumulative_PSF_counts} to derive constraints in the spatial extension of the emission.
Note that these are conservative lower limits, because higher S/N data may reveal that the emission extends farther out. 

Both sources show evidence of extended envelopes from various observations of gas and dust (e.g., Hartmann et al. 1993,
Grady et al. 2001, in the case of HD~100546; Doering et al. 2007, Martin-Za\"idi et al. 2010 in the case of HD~97048). 
However, the H$_2$ emission observed is not consistent with emission from an envelope.
In the case of HD~100546 the size of the envelope is $\sim$1000 AU from dust scattering imaging (Grady et al. 2001).
Because the line is spectrally resolved and its FWHM is much larger than the maximum expected Keplerian broadening at 1000 AU ($\sim$ 3 km s$^{-1}$),
envelope emission is ruled out. 
In addition,  if the emission would come from an envelope, its spatial extension would be much larger than measured in the PSF FWHM. 
A similar set of arguments holds for HD~97048. 
Furthermore, in this source the shape of the SPP and the line position-velocity diagram 
rules out emission from an envelope, and an H$_2$ line centered at the velocity of the star is inconsistent with the radial velocity shift (14.5 km s$^{-1}$, 
Martin-Za\"idi et al. 2010) observed in absorption lines (CH and CH$^+$) tracing HD~97048's envelope.

Finally, we note that the line flux measured in HD~97048 is one order of magnitude higher than reported by Bary et al. (2008) 
(8.6$\pm0.4\times10^{-15}$ erg s$^{-1}$ cm$^{-2}$) using Phoenix ($R\sim6$ km s$^{-1}$) on Gemini South.
The reason for this is unclear. 
It could be either due to observational reasons (difference in resolution, i.e., with higher spectral resolution,
it is easier to separate the line from the continuum; the use of AO also minimizes slit losses from the extended emission component) or  the line might be intrinsically variable.
Although our observations were not designed to test for variability, 
there is tentative evidence that the line is intrinsically variable.
We obtained  observations of HD~97048 with an identical telescope setup in two epochs separated by 15 months. 
In the first observation (April 2007)  the line has a flux of $10\pm3\times10^{-14}$ erg s$^{-1}$ cm$^{-2}$, and in the second epoch (July 2008)
the line flux is $7\pm2 \times10^{-14}$ erg s$^{-1}$ cm$^{-2}$ (see Fig. \ref{HD97048_average}). 
This change in the flux could be caused by variability of the line. However, 
note that the error bars of the two measurements overlap. Therefore, 
the change seen in the line fluxes
can also be due to the uncertainty on the determination of the continuum flux (30\%).

Comparing our measurements with those of Bary et al., 
we find that even allowing for a 30\% uncertainty in the continuum fluxes,
our line fluxes are still several times higher than those measured by  Bary et al. (2008).
This difference cannot only be attributed to uncertainties on the continuum flux.
Measurements especially designed for variability (e.g., spectroscopy and photometry measurements simultaneously obtained)
are needed to confirm the variability of the 1-0 S(1) H$_2$ line.
 
\subsection{Line ratios and excitation mechanism.}
The line ratios of the H$_2$ lines allow us to constrain the excitation mechanism and the temperature of the gas.
Because the 1-0 S(0) and 2-1 S(1) lines are not detected, we can only derive upper limits to the H$_2$ 1-0~S(0)/1-0~S(1) and 
2-1~S(1)/1-0~S(1) line ratios. 
For this the H$_2$ 1-0~S(0) and 2-1~S(1) line flux upper limits were divided by
the measured H$_2$ 1-0 S(1) line flux minus 30\% of uncertainty. 
In Table~\ref{line_ratios} we summarize our results. 
For comparison, we include in Table~\ref{line_ratios} the expected LTE emission line ratios for H$_2$ at diverse temperatures, 
H$_2$ emission excited by UV, X-ray, and shocks compiled by  Mouri (1994) (see their Fig. 3 and Table 2), 
X-ray excitation models of H$_2$ by  Tin\'e et al. (1997), and
models of H$_2$ emission from classical T Tauri star disks by  Nomura et al. (2007)
assuming dust grains with a spatially uniform distribution, dust well mixed with the gas,
and maximum grain radii $a_{\rm max}=10\,\mu$m.

We find that in the case of HD~100546,
the 1-0~S(0)/1-0~S(1) and the 2-1~S(1)/1-0~S(1) H$_2$ line ratio upper limits exclude 
pure radiative UV fluorescent H$_2$ emission from low-density gas (n~$\lesssim$~10$^4$ cm$^{-3}$).
However, the line ratio upper limits are too high to be able to distinguish between thermal, UV, X-ray, or shock emission.

In the case of HD~97048, the 1-0~S(0)/1-0~S(1) and  2-1~S(1)/1-0~S(1)  H$_2$ line ratios exclude pure radiative UV fluorescent H$_2$ emission in low-density gas.
Additionally, the 1-0~S(0)/1-0~S(1)  line ratio also excludes  emission by gas at T$<$2000 K,  
UV excited thermal and fluorescent H$_2$ emission from dense gas (n~$\gtrsim$~10$^4$ cm$^{-3}$), 
and shock excitation models.
Measured H$_2$ line ratio upper limits are consistent with the line ratios from thermal gas radiating at a temperature warmer than 2000 K
and/or from H$_2$ excited by X-rays.
We note that HD~97048 was detected in X-rays by ROSAT by Zinnecker \& Preibisch (1995). A short XMM-Newton observation (34 ksec; ObsID 0002740501, PI R. Neuh\"auser) serendipitously detected it again with a flux of about 8$\times 10^{-14}$ erg s$^{-1}$ cm$^{-2}$ (in the 0.3 -- 8 keV band; XMMXASSIST database, Ptak \& Griffiths 2003).

\begin{table}
\setlength{\tabcolsep}{1.1mm}
\caption{Measured upper limits to the H$_2$ emission line ratios and theoretical  line ratios expected for H$_2$ at LTE, and H$_2$ excited by UV,  X-ray, and  shocks
(Mouri 1994,  Tin\'e et al. 1997, Nomura et al. 2007).
}              
{
\centering                                      
\begin{tabular}{l c c   }          
\hline\hline 
\\ [-2.0ex]                     
 \multicolumn{1}{c}{} &{\large $\frac{1-0{\rm \,S(0)}}{1-0{\rm \,S(1)}}$ } & {\large $\frac{2-1{\rm \,S}(1)} {1-0{\rm \,S}(1)}$ }\\
						\\ [-2.0ex]  
\hline
\\ [-2.0ex] 
{\bf Observed}  \\
~~HD~97048   & {\bf $<	$0.20}& {\bf $<$0.22} \\
~~HD~100546 & {\bf $<$0.55} &{\bf  $<$0.39} \\[1mm]
{\bf LTE} \\  
~~$T =$  ~500 K & 0.44 & 1.8$\times10^{-5}$~~ \\ ~~$T =$ 1000 K & 0.27 & 0.005 \\
~~$T =$ 2000 K & 0.21 & 0.085 \\
~~$T =$ 3000 K & 0.18 & 0.21~~ \\[1mm]
{\bf UV pure radiative fluorescence} \tablefootmark{a} \\
~~$n_{T} = 10^2 - 10^4 $ cm$^{-3}$ \\
~~$~\chi= 1-10^4$
 & 0.38 -- 1.18  & 0.52 -- 0.58 \\[1mm]
{\bf UV thermal + fluorescence} \tablefootmark{b}\\
~~$n_{T} =10^5 - 10^6$ cm$^{-3}$ \\
~~$~\chi= 10^2 -10^4$ & 0.27 -- 0.34 & 0.0062 -- 0.025 \\[1mm]
{\bf X-ray} \\
~~Lepp \& McCray (1983) \tablefootmark{c}  & 0.23 & 0.010 \\
~~Draine \& Woods (1990) \tablefootmark{d}  & 0.21 - 0.20 & 0.097 -- 0.075 \\
~~Tin\'e et al. (1997) \tablefootmark{e} \\
~~~~{\tiny $T=500~~$ K, $n_{\rm H}=10^5-10^7$ cm$^{-3}$} & 0.28 -- 0.49 & 0.01 -- 0.35 \\
~~~~{\tiny $T=1000$ K, $n_{\rm H}=10^5-10^7$ cm$^{-3}$} & 0.28 -- 0.29 & 0.001 -- 0.006 \\
~~~~{\tiny $T=2000$ K, $n_{\rm H}=10^5-10^7$ cm$^{-3}$} & 0.21 -- 0.22 & 0.013 -- 0.082 \\[1mm]
{\bf Shock}  \tablefootmark{f} & 0.23 & 0.084 \\[1mm]
{\bf Nomura CTTS disk models} \tablefootmark{g} \\
~~X-ray irradiation & 0.23 & 0.06\\
~~UV	irradiation	     & 0.25 & 0.02 \\
~~X-ray+UV  irradiation         & 0.24 & 0.03\\
\hline
\label{line_ratios}
\end{tabular}
\flushleft
\tablefoot{\tiny
\tablefoottext{a}  Models of pure radiative UV fluorescent H$_2$ emission spectra produced in low-density (n~$\lesssim$~10$^4$ cm$^{-3}$) cold isothermal photodissociation regions from 
Black \& van Dishoeck (1987). 
\tablefoottext{b} Models of H$_2$ infrared emission spectra in dense (n~$\gtrsim$~10$^4$ cm$^{-3}$) static photodissociation regions exposed to UV radiation that both heats and excites the H$_2$ gas from Sternberg \& Dalgarno (1989).
\tablefoottext{a}\tablefoottext{b} Parameters: $n_{T}=$ the total density of hydrogen atoms and molecules; 
$\chi =$ UV-flux scaling relative to the interstellar radiation field;
\tablefoottext{c} X-ray heating models of Lepp \& McCray (1983) assume an X-ray luminosity in the 1-10 keV band of
10$^{35}$ ergs s$^{-1}$. Here are given the line ratios of their model b.
\tablefoottext{d}
X-ray excitation models of Draine \& Woods (1990) provide the expected H$_2$ line emissivity efficiencies
assuming a rate of absorption of X-ray energy $\gamma=2\times10^{-19}$ erg s$^{-1}$,
n$_{\rm H}=10^5$ cm$^{-3}$, and monochromatic X-rays of 100 eV. 
In the case of the H$_2$ 1-0 S(1) line, they obtain efficiencies of $9.1\times10^{-3}, 7.9\times10^{-3}$, and  $4.8\times10^{-3}$ 
for an X-ray energy absorbed per H nucleus of 1, 10, and 62.3 eV respectively.
In HD~97048 and HD~100546, we measured H$_2$ 1-0 S(1)  luminosities of $1.9\times10^{29}$ and $3.4\times10^{28}$ erg s$^{-1}$  respectively.
Using the Draine \& Woods (1990) H$_2$ 1-0 S(1) line efficiencies, 
these H$_2$ 1-0 S(1) luminosities would imply a $L_{\rm X}$ from $10^{30}$ to $10^{31}$ erg s$^{-1}$,
somewhat brighter than the $L_{\rm X}\sim 10^{29.5}$ measured in Herbig Ae stars (Telleschi et al. 2007). 
\tablefoottext{e} 
Tin\'e et al. (1997) models do not prescribe a specific X-ray input luminosity.
They assume electrons of 30 eV and use as input parameters the ionization rate $\zeta$ (10$^{-8}$ to 10$^{-17}$ s$^{-1}$), 
$n_{\rm H} (10 - 10^7$ cm$^{-3}$), and $T= 500, 1000, 2000$ K (see Tables 8 and 9 of Tin\'e et al. 1997).
These models include, in addition to the effects of X-rays, the effects of collision processes of H$_2$, with H$_2$, H, and He on the resulting H$_2$ emission spectrum.
\tablefoottext{f} Brand et al. (1989).
\tablefoottext{g} Nomura et al. (2007) CTTS models assume a $L_{\rm X}\sim10^{30}$ ergs s$^{-1}$ and
$L_{\rm FUV}\sim10^{31}$ ergs s$^{-1}$. Here are given the line ratios of the models with 
maximum grain radii $a_{\rm max}=10\,\mu$m. 
}
}
\end{table}    

\subsection{Line modeling with Keplerian flat-disk models}
\label{sec:model}
We modeled the H$_2$ line using a toy model that mimics emission from gas in a Keplerian orbit assuming a flat disk with known 
inclination and position angle (PA) (see Table \ref{model_results}).  
The intensity of the emission is designed to decrease as $I(R) \propto \left(R/R_\mathrm{in}\right)^{-\alpha}$, 
with $R_\mathrm{in}$ being the inner radius, and $R$ the radial distance from the star.  
We used a 2D grid from $R_{\rm in}$ to $R_{\rm out}$ and $\theta$ from 0 to 360.
For each grid point the emission is calculated by multiplying the intensity times, the solid angle times, 
a normalized Gaussian with FWHM of 3 km s$^{-1}$ (to simulate the spectral resolution) and center equal to the projected Keplerian velocity of the grid point.
The disk emission is convolved in the spatial direction with a 2D Gaussian with the same FWHM as the observed AO point-spread function. 
Then, we overlaid a slit on the simulated image and produced a  2D  spectrum.
The simulated 2D spectrum was rebinned in such a way that the pixel scale in the dispersion and the spatial direction is the same as in 
our CRIRES data. The rebinning was made conserving the total amount of flux in the 2D spectrum.
Finally, the flux of the simulated 2D spectrum was re-normalized such that the continuum has a value 1 in the 1D extracted spectrum,
and the amount of flux in the line inside $\pm$20 km s$^{-1}$ and $\pm$700 AU is equal to the flux of the line 
in the same region as the observed 2D spectrum. 
This re-normalization was made to enable us to compare directly the normalized 
2D observed spectrum with the 2D spectrum computed with the model.

Because the stellar parameters (mass, distance, inclination) are all constrained for HD~100546 and HD~97048 
in the literature (see Table \ref{model_results} and van der Plas 2009),  
the free parameters for the line modeling are $\alpha$, $R_\mathrm{in}$, and $R_\mathrm{out}$.   
The exponent $\alpha$ parametrizes the dependence of the intensity as a function of the radius;
it describes the combined effect of the variation of the density, temperature, and emissivity of the disk.
This assumption allows us to reduce the number of free parameters for the line modeling.
For example, if we assume that the surface density is constant, an $\alpha$ = 2 would reflect an intensity directly proportional to the stellar radiation field.

We compared the observed extracted 1D line profile and the 2D line spectrum (i.e., the 2D spectrum minus the continuum PSF) with 
their simulated counterparts for the parameter space 
$ 0.1~\mathrm{ AU}\leq R_\mathrm{in} \leq 450~ \mathrm{ AU} $,  
$ 10~\mathrm{ AU}\leq R_\mathrm{out} \leq 500~ \mathrm{ AU} $,
and $0 \leq \alpha \leq 5.5$, 
calculating for each model the reduced $\chi^2$ statistic:
\[
\chi^2_{\rm red} = \frac{1}{\sigma^2}\sum {\frac{(O - M)^2}{(N-4)}}~.
\]
Here, $O$ are the observed data points and $M$ the modeled values. 
The variance $\sigma^2$ was determined from the continuum emission between -50 and -20 km s$^{-1}$, and +20 and +50 km s$^{-1}$ 
at -300 to - 100 AU and +100 to +300 AU.
$\chi^2_{\rm red}$ was calculated over the data between -20 and +20 km s$^{-1}$.
$N$ is the number of valid data points between -20 and +20 km s$^{-1}$.
The number of degrees of freedom ($N-4$) is obtained because we use three free parameters for the fit  ($R_{\rm in}$, $R_{\rm out}$, $\alpha$),
and one degree of freedom is taken because  the simulated line is re-normalized to match the total 2D flux of the observed line.

\subsubsection{Modeling results}

\begin{figure*}
   \centering      
   \begin{tabular}{c}
\\[0.5cm]
\includegraphics[width=0.9\textwidth]{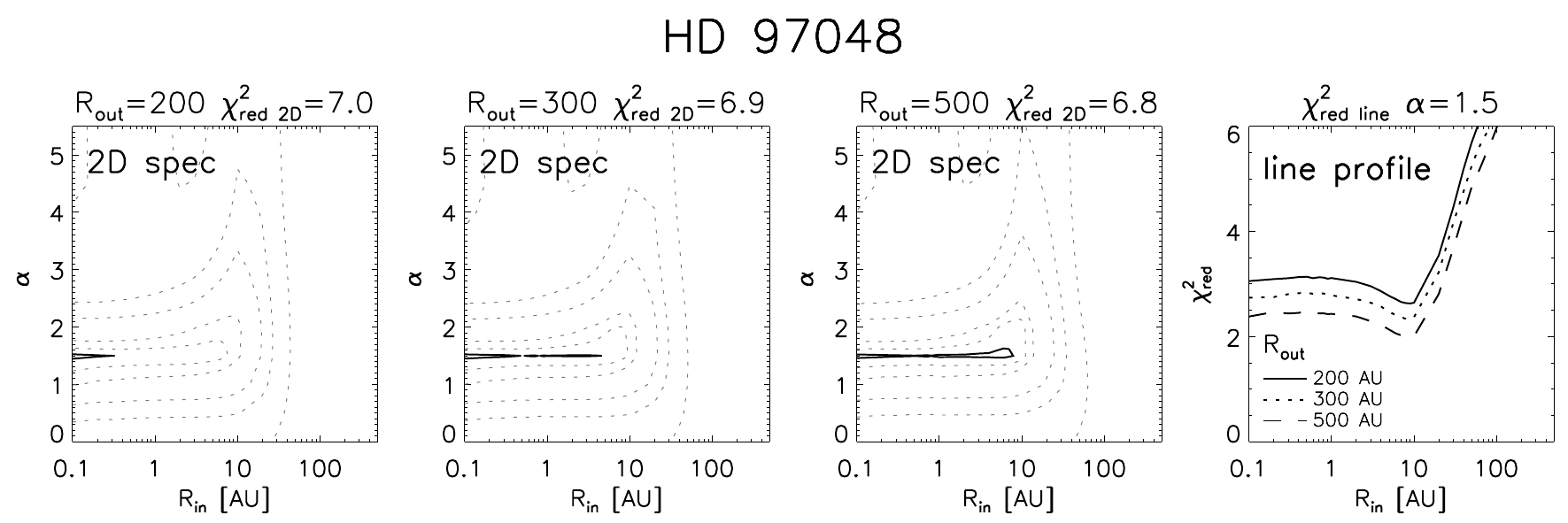}\\[1.5ex]
\includegraphics[width=0.9\textwidth]{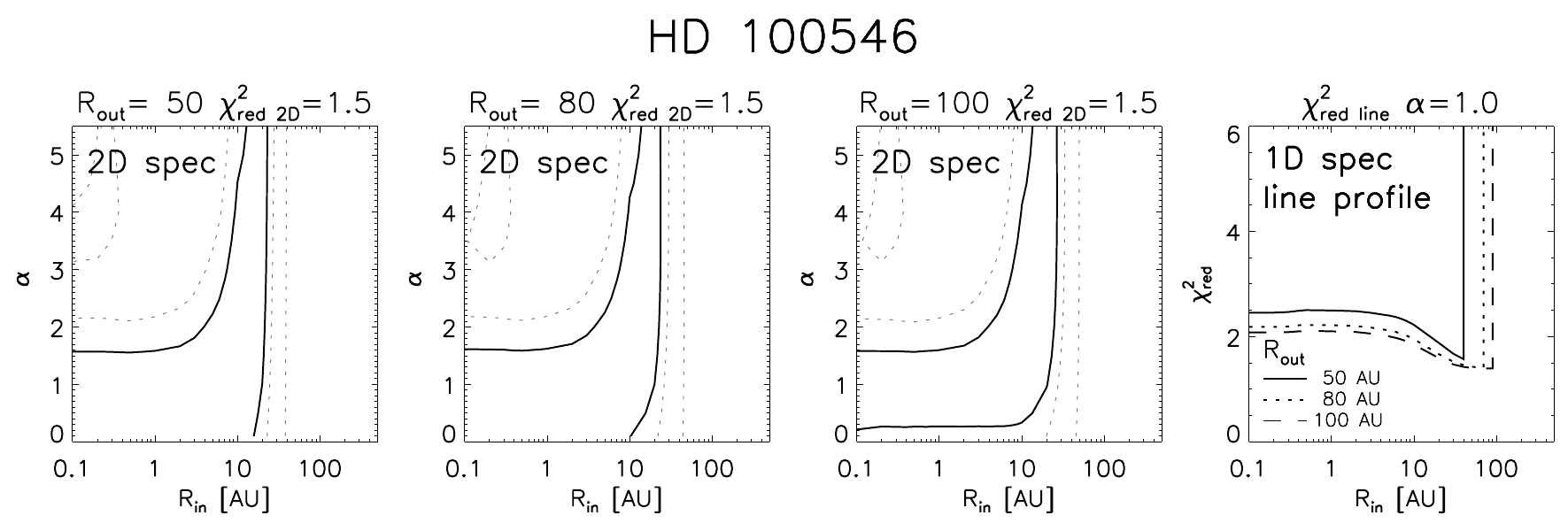}\\
\end{tabular}
\caption{ Reduced $\chi^2$ contours of the H$_2$ 2D and 1D line profile modeling for HD~97048 and HD~100546.
For each star the
{\it first three panels} display the reduced $\chi^2$ contours for the 2D spectrum as a function of  $\alpha$ and $R_{\rm in}$ for 
diverse values of $R_{\rm out}$ . The value of the minimum $\chi^2_{\rm red}$ is indicated in the title. The contour in solid black
is the contour of the minimum $\chi^2_{\rm red}$ plus 0.1. Subsequent contours in gray dotted lines are at the minimum $\chi^2_{\rm red}$ plus 0.5, 1, 3, 5, 10
in the case of HD~97048, and at the minimum plus 0.2 and 0.5 in the case of HD~100546.
The {\it rightmost panel} shows  the 1D line profile $\chi^2_{\rm red}$ as a function of $R_{\rm in}$ for several values of $R_{\rm out}$ for a fixed $\alpha$.  
For HD~97048 $\alpha$=1.5 is used, in the case of HD~100546 $\alpha$=1.0 is employed.
}
\label{fig_chisq_disk}
\end{figure*}

In Fig. \ref{fig_chisq_disk}
we show $\chi^2_{\rm red}$ contours for the continuum-subtracted 2D spectrum 
for diverse values of $R_ {\rm out}$ in the $\alpha$ vs $R_ {\rm in}$ parameter space,
and the $\chi^2_{\rm red}$ contours for the 1D extracted line profile as a function of $R_ {\rm in}$
for different values of $R_ {\rm out}$ for a fixed $\alpha$.
In the following paragraphs
we describe the modeling results in each source individually.
In Table \ref{model_results} we present a summary of the modeling of our data.

{\it HD~97048:} 
The 2D spectrum clearly shows an extended emission component extending at least up to
200 AU.
The $\chi^2_{\rm red}$ contours  for models with  $R_{\rm out} >$ 200 AU 
indicate that the solutions with minimum $\chi^2_{\rm red}$
converge to $R_ {\rm in}<$ 10 AU  and $\alpha$ varying 1.0 to 1.5 (see Fig.~\ref{fig_chisq_disk}).
The $\chi^2_{\rm red}$ decreases by increasing $R_ {\rm out}$.
This suggests that the emission may extend at least to 300-400 AU.
To constrain the multiple values of $R_ {\rm in}$ giving similar $\chi^2_{\rm red}$ in the 2D spectra,
we used the fit to the extracted 1D line profile.
By fixing $\alpha$ to 1.5, we calculated $\chi^2_{\rm red}$ for the line profile for diverse $R_{\rm out} >$ 200 AU.
Because the line observed is single peaked, 
solutions with radii larger than 3 AU provide a better fit.
The $\chi^2_{\rm red}$ of the line profile decreases up to $R_ {\rm in} \sim$ 8 - 9 AU,
then the $\chi^2_{\rm red}$ starts to rapidly increase (see Fig.~\ref{model_lines} top panel, upper right).
This suggests that the $R_ {\rm in}$ that best describes the line shape and the 2D spectrum simultaneously  
is $\sim$ 8 AU.
In summary, 
flat disk models indicate that the H$_2$ emission observed in HD~97048 
is most likely produced in the disk starting at $5<R_ {\rm in}<$10 AU,  
and that it extends to a $R_ {\rm out}$ of a few hundred AU.
In Fig.~\ref{model_lines}, we present the line profile, FWHM, SPP, 2D spectrum, 2D model spectrum and their difference
for a disk model with $R_ {\rm in}$=9 AU, $R_ {\rm out}$=400 AU and $\alpha$=1.5. 

{\it HD~100546:}  
The detected H$_2$ 1-0 S(1) in HD~100546 is weak and the amount of counts in the 2D spectrum 
is low. Therefore, the direct comparison to 2D model data is challenging.
The first observational fact is that the line observed is single-peaked
and that the line wings are $<$ 10 km s$^{-1}$.
This is better described  by disk models with emission at large inner radii $R_ {\rm in}$ 30 -100 AU (see lower right panel of Fig.~\ref{fig_chisq_disk}).
In contrast, the position velocity diagram of the 2D spectrum displays the emission relatively
concentrated at $R_ {\rm out}<$50 AU, which favors disk solutions at $R_ {\rm in}<20$  AU (see lower left panels of Fig.~\ref{fig_chisq_disk}).
Thus, the best models are a compromise between reproducing the single peak and the lack of high velocity wings
in the line profile (i.e., favoring a large $R_ {\rm in}$), while keeping the emission not too extended to fit the 2D spectrum.
In the lower-left panels of Fig.~\ref{fig_chisq_disk}, 
we present the $\chi^2_{\rm red}$ contours for flat disk models with $R_{\rm out}=$ 50, 80 and 100 AU.
We observe that there is a large family of solutions with similar $\chi^2_{\rm red}$.
If we assume $\alpha$ to be larger than 1.5, the solutions converge to a disk with $R_{\rm in}$ between 5 and 20 AU.
If $\alpha$ is lower than 1.5, we can only say that $R_{\rm in}$ should be smaller than 20.
To break this degeneracy, the 1D line profile is useful. 
In the lower-right panel of  Fig.~\ref{fig_chisq_disk} 
we display $\chi^2_{\rm red}$ of the 1D line profile as a function of $R_{\rm in}$ for diverse values of $R_{\rm out}$.
This plot shows that the $\chi^2_{\rm red}$ is almost constant up to 10 AU, 
then decreases reaching its minimum between 30-100 AU (depending of $R_ {\rm out}$),
and then increases rapidly. 
This plot suggests that $R_{\rm in}>$ 10 AU, and that for a given $\alpha$ the best combined fit of line profile and 2D spectrum
is given by the maximum  $R_{\rm in}$ inside the contour of the minimum $\chi^2_{\rm red}$ in the 2D spectrum.
In summary, 
flat disk models indicate that the H$_2$ emission observed in HD~100546
is most likely produced in the disk starting at $10<R_ {\rm in}<20$ AU  and that the emission extends at least to 50 AU.
In Fig.~\ref{model_lines} we present the line profile, FWHM, SPP, 2D spectrum, 2D model spectrum, and their difference
for a disk model with $R_ {\rm in}$=20 AU, $R_ {\rm out}$=80 AU and $\alpha$=1.0. 

The flat disk-models used here are ``toy models" and  are oversimplifications of the real structure of the disks,
which are indeed flared-disks with (at least in the case of HD~100546) a complex inner disk structure in the dust (see Section 4.2).
In a flared-disk geometry the outer regions of the disk are more exposed to radiation than in a flat-disk geometry.
Therefore, it is expected that the effect of the flared geometry would be to displace the $\chi^2_{\rm red}$ contours 
towards lower values of alpha to account for the increment in the flux contributions from the outer disk.
Since the inner radius is less affected by the flaring, the inner radius of the emission 
of a  flared disk will not change considerably with respect to the inner radius of a flat disk.
Note in particular that the lowest values of $\chi^2_{\rm red}$ are $>$1.5, 
which formally means that the models do not  fit the data correctly.
Nevertheless,
the main message that the flat disk models are telling us is that 
most of the H$_2$ emission observed is produced at $R>>$5 AU.
This is an important message because, in the context of passive disk models with gas and dust in thermal equilibrium
developed for Herbig Ae/Be stars (e.g., Dullemond et al. 2001), the H$_2$ gas at temperatures of few thousand K,
which is necessary for emitting the near-IR lines, is expected to be located in the inner parts of the disk only up to a few AU. 
We will discuss this in detail in Section 4. 

\section{Discussion}
\subsection{H$_2$, CO, and {\rm[}\ion{O}{i}{\rm ]} emitting regions.}
As a summary of our observations and modeling,
we can conclude that in both sources
the observed near-IR H$_2$ emission appears to be produced at distances
larger than 5 AU.
In the case of HD~97048 the emission extends at least up to 200 AU and 
in the case of HD~100546 the emission extends at least up to 50 AU.  
In both targets, 
two other gas tracers, the [\ion{O}{i}] line at 6300 \AA\ 
and the CO ro-vibrational emission band at 4.7 $\mu$m, 
have been detected (e.g. van der Plas et al. 2009, Acke et al. 2005).
In addition, in the case of HD~97048 the H$_2$ 0-0 S(1) line at 17~$\mu$m has been observed (Martin-Za\"idi et al. 2007).
How does the near-IR H$_2$ emitting region compare with these other tracers?

Seeing-limited observations of the 0-0 S(1) H$_2$ line at 17 $\mu$m by Martin-Za\"idi et al. (2007) 
set an upper limit to the extension of the emission of 35 AU. 
Because the 17 $\mu$m line is spectrally unresolved in the Martin-Za\"idi et al.~~spectra  (FWHM$\sim$30 \kms),
the profile cannot be directly compared with our CRIRES 1-0 S(1) profiles.
It can only be said that the inner radius producing the 17 $\mu$m line should be larger than a few AU, 
otherwise the line would have been spectrally resolved.
Given that the 1-0 S(1) H$_2$ emission at 2 $\mu$m 
is located farther out in the disk (at least up to 200 AU),
the two data sets clearly indicate that the emitting regions of the 2 and 17 $\mu$m line are different,
with the 1-0 S(1) line excited farther out than the 0-0 S(1) line.
Note that in the case of thermal disk emission, the contrary is expected, 
namely that the 17 $\mu$m line is produced farther out because the temperature decreases with the radius.
This indicates that the excitation mechanism of the two lines is different,
or at least very inefficient at producing detectable levels of 0-0 S(1) at large distances.
Models of H$_2$ emission from disks around T Tauri stars (e.g. Nomura et al. 2007),
predicted 0-0 S(1)/ 1-0 S(1) line ratios raging from 0.13 to 40 for fixed grain sizes (10 $\mu$m to 10 cm) and  
30 to 70 for models including grain coagulation and settling.
The measured line ratio from our and Martin-Za\"idi's observations is 
0.25$^{+0.2}_{-0.1}$ assuming a 30\% uncertainty in the continuum.
This ratio is similar to that suggested for the UV  and/or X-ray excitation models with grains of 10 $\mu$m size.
It does not agree with the ratio from models with larger grains or those implementing dust coagulation and settling.

\begin{table}
\caption{Modeling results summary.}              
\flushleft                                  
{\normalsize
\begin{tabular}{c |c l c c c l}          
\hline
\hline
 & & &$R_{\rm in}$ & $R_{\rm out}$ &  \\
Star & Species & Model &[AU] & [AU] & $\alpha$ \\ 
\hline
\\[-1.5ex]
{\bf HD~97048}   	& H$_2$ & Flat disk  & 5-10 & $>$200 & 1.5 \\
$i=47^\circ$			& CO $^a$  & Flat disk  & 11  & $>$50  & 2.5 \\
PA=175$^{\circ}~$\tablefootmark{c}	& [\ion{O}{i}] \tablefootmark{b} & Flared &  0.8 & $\sim$50  \\	[2mm]							   									 
\hline
\\ [-1.5ex]
{\bf HD~100546}  & H$_2$ & Flat disk  & 10-20 & $>$50 & ind. \\
$i=43^\circ$			& CO \tablefootmark{a}  & Flat disk     & 8  & $>$50  & 2.5 \\
PA=160$^\circ$	& [\ion{O}{i}] \tablefootmark{b}& Flared  &  0.8 & $\sim$50  & ... \\[2mm]
\hline						 	
\end{tabular}
\flushleft
\tablefoot{
\tablefoottext{a} van der Plas (2009); 
\tablefoottext{b} Acke et al. (2005); 
\tablefoottext{c} The position angle (PA) of the major axis of the disk around HD~97048 on the sky is 175$\pm$1$^\circ$ E of N, derived from the 8.6\,$\mu$m image of Lagage et al. (2006). 
This value agrees with the PA previously determined from spectro-astrometry of the [\ion{O}{i}] 6300\,\r{A} line (160$\pm$19$^\circ$, Acke et al. 2005). 
}
\label{model_results}
}
\end{table}

\begin{figure}
\centering                  
\includegraphics[width=0.47\textwidth]{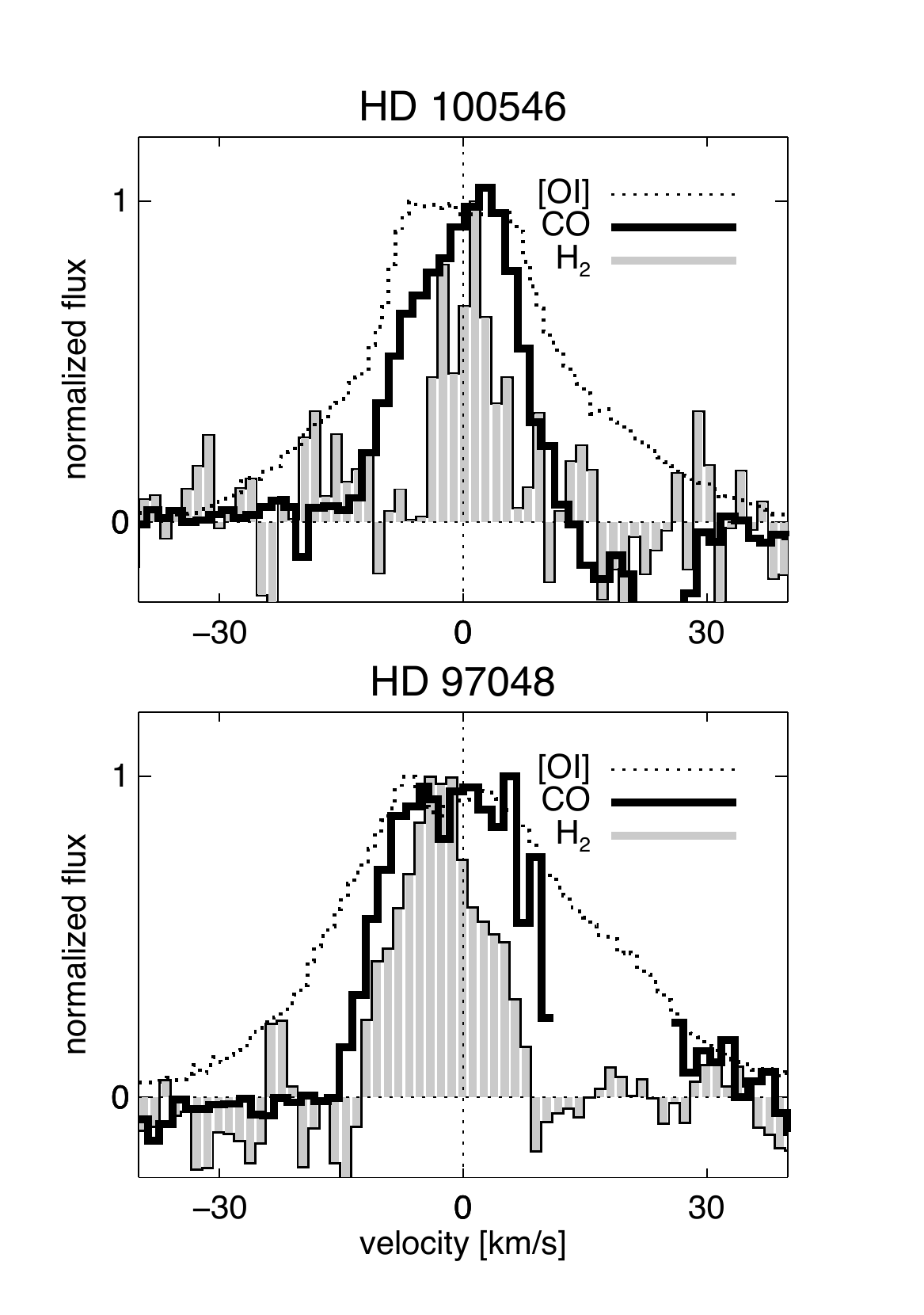}
\caption{Observed line profiles of  H$_2$ 1-0 S(1) emission (in gray, histogram), combined $\upsilon=1-0$ CO emission, and [\ion{O}{i}] emission for
HD~100546 (top) and HD~97048 (bottom). All three spectra have similar spectral resolution.}
 \label{profile_gas_tracers}
\end{figure}

In the case of the [\ion{O}{i}] line at 6300 \AA~ and the CO ro-vibrational spectra at 4.7 $\mu$m,
the spectral resolution of the data is similar to that of our observations. 
Therefore,  the line profiles can be used to compare the different emitting regions.
In Fig. \ref{profile_gas_tracers}
we plot the [\ion{O}{i}] line spectrum, 
the composite $\upsilon=1-0$ CO ro-vibrational spectrum,
and our H$_2$ 1-0 S(1) line observations.
To facilitate the comparison, 
the data are presented in such way that the lines are centered at the rest velocity,
the continuum is set to zero, and the peak of the 
lines is set to 1.
All spectra have similar spectral resolution.
In this way we can compare the width and the high-velocity wings of the different transitions.
Additionally, we include in Table \ref{model_results} the results of kinematic modeling of the $\upsilon=1-0$ CO (van der Plas et al. 2009) 
and [OI] (Acke et al. 2005) line profiles.

As a general trend in both sources,
we observe the following characteristics:
{\it (i)} The [\ion{O}{i}] line is the broadest line and has velocity wings extending to $\pm$50 km s$^{-1}$;
{\it (ii)} The CO line profile is narrower than the [\ion{O}{i}] line and broader than the H$_2$ line;
{\it (iii)} The [\ion{O}{i}] profiles are broad and double-peaked;
{\it (iv)} The CO lines are broad, and the case of HD~97048 they are consistent with a double-peaked profile
(van der Plas (2010) detected the $\upsilon=2-1, 3-2, {\rm and}~4-3$ CO transitions in HD~100546 and HD~97048,
 the observed line profiles are consistent with double-peaked profiles).
 In contrast, the H$_2$ line is single-peaked.
Assuming that we are observing disk emission,
these characteristics indicate that each gas tracer is produced in a different radial region of the disk.
The [\ion{O}{i}] line is produced the closest to the star,  
it is followed by the CO emission at a larger distance,
 and finally the H$_2$ line farther out in the disk.

\subsection{H$_2$ near-IR emission and observations of dust in the inner disk}

HD~97048 and HD~100546 have been extensively studied in dust continuum emission,
including studies with ground-based infrared interferometers in the case of HD~100546 (e.g., Benisty et al. 2010). 
How does the H$_2$ near-IR line-emitting region
compare with  inner disk structure deduced from dust observations?
 
In the case of HD~100546, 
recent simultaneous modeling of the spectral energy distribution (SED) and near-IR interferometry 
(Benisty 2010, Tatulli et al. 2011)
suggested a dust disk structure consisting
of a tenuous inner disk located around 0.25 AU to 4 AU, 
followed by a gap devoid of dust,
and a massive outer disk starting at 9 to 17 AU that extends up to a few hundred AU.
Comparing this disk dust structure and the H$_2$ near-IR emitting region,
we find  that the inner radius of the region emitting the H$_2$ line
and the inner radius of the massive outer dusty disk 
are remarkably similar. 
If the inner radius of H$_2$ near-IR emission
and the beginning of the outer thick disk are indeed the same,
this would suggest that the presence of the gap in the dust is favoring the excitation
of H$_2$ at large distances from the star. 
However,  it still remains to be explained in this scenario,  
why [OI] and CO emission are present at radii smaller than H$_2$.

In the case of HD~97048, no near-IR interferometry observations are reported in the literature. 
The most comprehensive study of the dusty disk structure is from Doucet et al. (2007),
who simultaneously modeled the SED, mid-IR spectra and spatially resolved mid-IR imaging in the continuum and 
PAH feature at 11.3 $\mu$m. Those authors suggest a continuos flared disk model with a puffed-up inner rim at radius 0.4 AU, 
outer radius 370 AU, and flaring index 1.26. 
The extension of the dust emission down to a fraction of AU combined with the detection of gas tracers
such as the [OI] line in HD~97048 also down to a fraction of AU (Acke et al. 2005),
contrasts with the H$_2$ near-IR emitting region at
a radius larger than 5 AU found in our observations. 
The reason for this discrepancy is unclear. 

\subsection{H$_2$ near-IR emission in the context of diverse disk structure models}
In summary, 
we find that in HD~100546 and HD~97048,
H$_2$ near-IR emission appears only to be present at large radii.
But, as discussed, 
within these radii material is still present within the disk, as 
shown by the presence of warm dust  (e.g. Bouwman et al. 2003; Tatulli 2011; Doucet 2007), 
the kinematics of the [\ion{O}{i}] emission line at 6300\,\AA\ 
(Acke \& van den Ancker 2006) and the CO ro-vibrational band at 4.7 $\mu$m (van der Plas et al. 2009). 
The situation as seen in the H$_2$ 1-0 S(1) line reported here is somehow
analogous to what is seen in CO around these and other Herbig stars
(Brittain et al. 2009; van der Plas et al. 2009; 2010), where there also appears 
to be a deficit of CO emission coming from the inner regions of the disk.
Van der Plas et al. (2009) have argued that this may imply 
that CO is efficiently depleted close to the star.  
However, 
it is unlikely that this explanation will also hold for H$_2$, given its much greater abundance;
one would only expect H$_2$ to be depleted in a disk that is gas poor overall.
Moreover, the detection of [O\,{\sc i}] emission close to the star suggests
that the disks around HD~97048 and HD~100546 are still gas-rich, 
which eliminates this possibility. 
To address this and explain the numerous non-detections of H$_2$ near-IR emission,
we discuss  below our results in the frame of
passive disks models with  thermally coupled gas and dust,
passive disks models with  thermally decoupled gas and dust, 
and disk photoevaporation.

\subsubsection{H$_2$ near-IR emission in the context of passive disks with  thermally coupled gas and dust}
Standard passive disks models of Herbig Ae/Be disks,
i.e., models  in which gas and dust are thermally coupled in the disk surface layer (e.g., Dullemond et al. 2001),
show that in the outer regions of the disk ($R>5$ AU) the temperatures in the disk 
are too low to emit efficiently  in the detected H$_2$ 1-0 S(1) line, 
which requires gas of a few thousand degrees to be excited.
Employing Chiang and Goldreich (1997) Herbig Ae two-layer disk models
(using the implementation by 
Dullemond et al. 2001 described in Carmona et al. 2008b),
we calculated the amount of  gas in the surface layer at T$_{\rm surf}>$1000 K 
for disks of mass 10, 40 and 100 M$_{\rm J}$
(see Figs. 4 and 5 of Carmona et al. 2008b).
We found that the mass of gas T$_{\rm surf}>$1000 K is $\sim$0.01 M$_{\rm moon}$.
Therefore, in the inner disk, where temperatures are sufficiently high to thermally (i.e., by collisions with dust)
excite the H$_2$ lines,  the amount of gas in the optically thin surface layer of the disk is small,
quenching the formation of the near-infrared H$_2$ lines.
Because CRIRES observations are typically sensitive from 0.1 to 1 M$_{\rm moon}$ of H$_2$ gas at T$>$1000 K (Carmona et al. 2007),
our first conclusion is that the lack of near-IR H$_2$ emission in the 13 sources with non-detections fully agrees with what would be expected 
from passive Herbig Ae/Be disks with  thermally coupled gas and dust.

The puzzling new observational result presented in this paper is that H$_2$ emission is present at radii $>>$5 AU in the disks around
the Herbig Ae/Be stars HD~97048 and HD~100546 \footnote{Detections of near-IR H$_2$ emission displaying an extended ring structure at 73-219 AU away from the star
have been reported toward the weak-lined T Tauri star  (WTTS) DoAr 21 (O. Panic 2009 PhD thesis). 
DoAr 21 is one of the brightest X-ray WTTS (Neuh\"auser et al. 1994). }. 
Because this is not expected in the context of passive disks with  thermally coupled gas and dust,
we conclude that at least in these two objects,
the disk structure is not accurately described by simple passive disk models.
Either {\it (i)} at large radii molecular gas and dust in the disk surface layer have departed 
from thermal coupling  (i.e., T$_{\rm gas} >$ T$_{\rm dust}$) 
and/or  {\it (ii)} the disk atmosphere must be much more extended at large radii, for example in the form a disk wind caused by photoevaporation.
We note that there is observational evidence that dust and gas thermally decouple in the surface layers of Herbig Ae/Be disks.
Fedele et al. (2008) studied the dust (traced by the 10 $\mu$m feature) and  the gas (traced by the [\ion{O}{i}] line at 6300\,\AA) in the disks of a sample of Herbig Ae/Be stars.
They found a difference in the gas and dust vertical structure beyond 2 AU in the Herbig Ae/Be star HD~101412, thereby providing evidence  
of gas-dust decoupling in a protoplanetary disk atmosphere.

\subsubsection{H$_2$ near-IR emission in the context of passive disks with thermally decoupled gas and dust}
Models of Herbig Ae/Be hydrostatic disks
allowing the decoupling of gas and dust in the disk surface layers 
and including additional heating mechanisms of the gas have been recently developed
(e.g., {\sc ProDiMo}, Woitke et  al. 2009).
These types of models can  provide a hint at the possible origin of near-IR H$_2$ emission observed
in HD~100546 and HD~97048.
In these models, the temperature of the gas in the surface layer of the disk can be much higher than the
gas temperature in the passive disk models where gas and dust are thermally coupled. 
The thermally decoupled gas reaches temperatures up to a few thousand K
at distances up to 10-20 AU in Herbig Ae/Be flared disks. 
In self-shadowed disks, the temperature is also high,
but is constrained to a radial extent of $<$ 5 AU (van der Plas, 2010).

In the {\sc ProDiMo} model,
the transition between H and H$_2$ in the disk is determined by the balance
between the formation rate of H$_2$ on grains and the dissociation of H$_2$ by
the stellar UV field. 
At high temperatures H$_2$ has a low formation efficiency. 
The sticking of H on grains to form H$_2$ is high,
but the lifetime of H on the surfaces is very
short owing to thermal desorption if the grain temperature is also high. 
As soon as the H$_2$ formation rate can balance the H$_2$ dissociation,
abundances start to build quickly. 
Otherwise abundances stay
low (on the order of 10$^{-5}$). 
In the models discussed by van der Plas (2010), 
the H/H$_2$ transition occurs at the location where the gas temperature drops below
$<$ 1000 K, and/or where the UV opacity is high enough to
protect the H$_2$ against photo-dissociation, or around ~10 AU.

Thermally decoupled gas and dust passive disk models would naturally explain why the H$_2$ emission 
is observed in flaring disks but not in self-shadowed disks, and explain the numerous non-detections.
However, it is not clear why then H$_2$ near-IR emission is not present in all the sources with flared disks in our sample.
One possible explanation is that HD~97048 and HD~100546 are flared disks with the earliest 
spectral types (A0V and B9V respectively) of the flaring disks that we have observed.
The earlier the spectral type, the more stellar radiation is available to heat the upper-disk layers. 
An additional challenge for the hydrostatic thermally decoupled gas-dust models is that
at least in the case of HD~97048, 
the line ratio limit of the detected 1-0 S(1) line to the (undetected) 1-0 S(0) line sets the gas temperature around 2000 K.
Because gas at $T\sim1000$ K is more abundant than gas at $T\sim 2000$ K,  
copious amounts of  H$_2$ emission from gas  at $T\sim1000$ K would be expected,
and gas at this temperature is not observed.
This suggests that in addition to decoupling of the dust and gas some change in the disk structure 
at radii $>$5 AU is also necessary to explain our observations. 
  
\subsubsection{An extended disk atmosphere caused by photoevaporation~?}  
It is interesting to consider the effect of photoevaporation in the context of the possible presence of an extended disk atmosphere at large radii
in the cases of HD~97048 and HD~100546.
H$_2$ near-IR emission is detected at a velocity consistent with the rest velocity of the stars. 
However, owing to uncertainties on the radial velocity of the sources,
the possibility that the lines have a blue shift of a few km s$^{-1}$ cannot be completely excluded.
A small blue shift would be consistent with an origin of the emission in a photoevaporative disk wind.

Photoevaporation of disks by ionizing photons is a conceptually simple process in which energetic radiation 
(far-UV (FUV), 6 eV $<hv<$ 13.6 eV; 
extreme-UV (EUV), 13.6 eV $<hv<$ 0.1 keV ;
and  X-rays $hv>$ 0.1 keV) from the central star
heats hydrogen at the disk surface, producing a hot ionized layer.  
Near the star this ionized zone above the disk is expected to be almost static.  
However, at large radii from the star the thermal layer will be unbound, powering a thermally 
driven disk wind (Shu et al. 1993; Hollenbach et al. 1994).  
The critical radius at which the upper layer of the disk will become gravitationally unbound is on the order of 10~AU in the 
case of T~Tauri stars and scales linearly with the mass of the central star (Font et al. 2004).  
This critical radius is compatible with the inner radii for  the H$_2$ 1-0 S(1) emission observed in HD~97048 and HD~100546.

The H$_2$ 1-0 S(1) emission observed in both HD~97048 and HD~100546 could 
thus be due to the presence of an extended disk atmosphere (i.e., disk wind)
at large radii owing to photoevaporation.  

In the case of HD~97048, 
this low velocity disk wind could also be responsible for the pure rotational H$_2$ emission at 17 $\mu$m  
that has been previously detected (Martin-Za\"idi et al. 2007).
A critical radius larger than 10 AU could explain why the 17 $\mu$m line is observed spectrally unresolved.
However, it remains to be understood, why the emission at 17 $\mu$m appears to be confined at R$<$ 35 AU,
while the emission at 2 $\mu$m micron extends up to hundreds of AU. 
One possibility is variability of the 17 $\mu$m  and  2 $\mu$m line because the two data sets were not obtained simultaneously.

In the case of HD~100546, 
Lecavelier des Etangs et al. (2003)  and Martin-Za\"idi et al. (2008a) reported 
H$_2$ in absorption in the FUV domain with {\it FUSE} \footnote{Far Ultraviolet Spectroscopic Explorer.}
at a radial velocity consistent with a circumstellar origin at the resolution of {\it FUSE} (R$\sim$15000).
Warm and hot gas at 760 K and 1500 K kinetic temperature were measured and revealed
excitation conditions of H$_2$ clearly different from those observed in the interstellar medium.
Because the {\it FUSE} line of sight does not pass through HD~100546 disk, 
those authors concluded that the H$_2$ observed by {\it FUSE} is not located in the disk
and suggested that the H$_2$ absorption might be produced by an FUV-driven photoevaporative wind 
from the outer parts of the disk.
An interesting possibility is that precisely this hot H$_2$ at T$>$1000 K from a photoevaporative wind 
seen by {\it FUSE} is being traced by the near-IR H$_2$ emission lines.

Models of EUV photoevaporation of disks around T Tauri stars
predict for disks with inclination $\sim$45$^\circ$ single-peaked emission lines of ionized gas
with blueshifts ranging from -12 to -7 km s$^{-1}$ and line FWHM ranging from 26 to 30 km s$^{-1}$
(e.g., [\ion{N}{ii}],  [\ion{S}{ii}] Font et al. 2004; [\ion{Ne}{ii}] Alexander et al. 2006, 2008).
These values are somewhat higher than the line center and line FWHM of the H$_2$ lines observed in HD~100546 and HD~97048
(see Table \ref{line_properties}), even taking into account the uncertainty on the radial velocity. 
If the H$_2$ emission observed is linked to an EUV photoevaporative wind, 
the velocity of the outflowing gas should be lower than a few km s$^{-1}$ to be consistent with our observations. 
The smaller line FWHM observed may be explained by the fact 
that the H$_2$ line is dominated by contributions at large radii.

In the context of  photoevaporative disk winds driven by FUV or X-rays (e.g., Gorti \& Hollenbach 2009, Ercolano \& Owen 2010),
there might be a connection between the presence of extended H$_2$ near-IR emission
and the rich and spatially extended PAH spectrum observed in HD~97048 and HD~100546 (van Boekel et al. 2004, Habart et al. 2006).
In contrast to EUV-driven disk winds, 
FUV and X-ray driven disk winds are predominantly denser, cooler, mainly neutral, and may drag along small grains.
	
The photoevaporative wind scenario does immediately raise the question why the H$_2$ lines we observe here have 
only been detected in HD~97048 and HD~100546 and not in the other 13 Herbig Ae/Be 
stars in which we searched for the emission.
In the case of objects with self-shadowed disks,   
the temperatures in the disk surface layer may not be high enough to launch a disk wind.
In the case of  flared disks, 
HD~97048  and HD~100546 are of relatively early spectral types (A0 and B9, respectively),
so photoevaporation or direct disk heating from stellar UV radiation could be more prominent in these stars. 
In addition, HD~97048 and HD~100546 are the Herbig stars with the most extreme 
UV fluorescence detected in CO ro-vibrational emission at 4.7 $\mu$m (see van der Plas et al. 2009, 2010)
and have the strongest PAH features at 8.6 and 11 $\mu$m (Acke \& van den Ancker 2004).
The two other sources with flared disks in our sample, HD~169142 and HD~179218, have slightly different physical properties than
HD~97048 and HD~100546.
HD~169142 has a spectral type A5. 
Because it is fainter than HD~97048 and HD~100546, 
the gas in the disk may be not sufficiently excited to produce detectable levels of H$_2$ near-IR emission. 
HD~179218 has a much higher luminosity than the other Herbig stars in the sample, but happens to have similar  T$_{\rm eff}$ as HD~100546. 
The higher luminosity might cause photodissociation of H$_2$ to larger distances from the star.
In this object van der Plas (2010) have detected warm CO at 4.7 $\mu$m likely excited trough UV fluorescence,
albeit much weaker than in HD~97048 and HD~100546.
The fact that HD~179218 has a distance of 240 pc may explain why the CO emission is weaker and the H$_2$ emission 
undetected. 
Alternatively, the photons responsible of photoevaporation could be driven by accretion, which may be somewhat above 
average in HD~97048 and HD~100546 (van den Ancker 2005).  

 \section{Conclusions}
We presented CRIRES high-spectral resolution observations of H$_2$ ro-vibrational emission
at 2 $\mu$m toward a sample of 15 Herbig Ae/Be stars. 
We detected 1-0 S(1) H$_2$ emission from 
HD~97048 and HD~100546 at a velocity consistent with the rest velocity of the stars;
in the other sample stars the emission is absent.
None of the sources displays the H$_2$ 1-0 S(0) or the 2-1 S(1) line.
In the case of HD~97048 the emission extends at least up to 200 AU. 
The spatial peak position (SPP) and the position-velocity diagram
show that  the emission at negative velocities has a positive spatial offset and that the emission at positive velocities has a negative spatial offset.
This provides evidence that the emission is observed from rotating material, most likely linked to the circumstellar disk.
The H$_2$ 1-0 S(1) flux measured in HD~97048 is one order of magnitude stronger than the line flux reported by Bary et al. (2008)
in observations carried out in 2003. 
Because the difference in the line fluxes between our and Bary et al.'s (2008) measurements is 
much larger than the uncertainty on the continuum flux, this suggests line variability.
In the case of HD~100546, the emission extends at least up to 50 AU.
The SPP is compatible with the signature of material in rotation; however,
observations with higher S/N are required to firmly establish this.

For HD~97048 and HD~100546,
we determined upper limits to the 1-0 S(0)/1-0 S(1) and 2-1 S(1)/1-0 S(1) line ratios 
and  constrained the excitation mechanisms.
In both sources pure radiative UV fluorescent H$_2$ emission from low-density gas (n~$\lesssim$~10$^4$ cm$^{-3}$) is excluded.
In the case of HD~100546 the line ratios upper limits did not permit us to distinguish  between
UV (thermal and fluorescent) excited H$_2$ emission in dense gas  (n~$\gtrsim$~10$^4$ cm$^{-3}$), 
X-ray, shocks, or thermally excited H$_2$ emission.
In the case of HD~97048, 
the line ratios upper limits are consistent with thermal emission at $T>$2000~K and/or H$_2$ gas excited by X-rays.

Assuming a flat disk in Keplerian rotation, 
we modeled the line profiles, 
the PSF FWHM, the SPP and the 2D spectrum
of the 1-0 S(1) H$_2$ line observed in HD~97048 and HD~100546. 
In both cases the models indicate that the emission arises at large ($>>$5 AU) radii.
In the case of HD~97048, the models suggest that the emission is produced starting from a $R_{\rm in}$ of 5-10 AU  and extends at least to 200-300 AU. 
Solutions converge to an $\alpha\sim$1.5.
In the case of HD~100546, the models suggest that the emission is produced starting from a $R_{\rm in}$ of 10-20 AU and extends at least to 50 AU.

In conclusion: 

{\it a)} The non-detections of near-IR H$_2$ can be explained {\it (i)} as a natural consequence of a passive disk structure
in which the gas and dust are thermally coupled (i.e., the amount of optically thin hot H$_2$ gas in the inner disk is too low to produce 
detectable levels of H$_2$ emission). Or  {\it (ii)}  in the context of a passive disk with thermally decoupled dust and gas; 
here the H$_2$ close to the star is photo-dissociated, and temperatures in the outer disks of self-shadowed disks are too low to produce H$_2$ emission.

{\it b)} Most of the H$_2$ emission observed in HD~100546 and HD~97048 is produced at $R>>5$ AU and extends at least to $R>50-200$ AU.
This location is much farther out in the disk than the radii up to a few AU where temperatures in the disk are expected to reach the 1000-3000 K that is needed to 
excite the observed H$_2$ lines thermally in standard passive disk models with thermally coupled gas and dust.
The emission can be explained by {\it (i)} disk models with thermally decoupled gas and dust that allow for high gas temperatures at large radii in flaring disks. 
At small radii, H$_2$ is photo-dissociated by the strong stellar UV field and H$_2$ formation efficiencies are low. 
After $\sim$ 10 AU, H$_2$ formation becomes more efficient at balancing  photo dissociation, thus leading to detectable H$_2$ gas in the disk atmosphere; 
or {\it (ii)} by  an extended disk atmosphere owing to a photoevaporative disk wind.
More detailed modeling would be of great help to understand the origin of the H$_2$ near-IR emission observed  in HD~100546 and HD~97048.

\acknowledgements{The authors wish to thank the anonymous referee and the A\&A editor M. Walmsley for the suggestions to the paper.  
A.C and M.A acknowledge support from Swiss National Science Foundation grants (PP002-110504 + PP002--130188).
We thank the Paranal nighttime astronomer E. Valenti for her cheerful assistance performing the observations, and
A. Mueller for discussions concerning the radial velocities of HD~100546 and HD~97048.
A.C thanks C. Dullemond and B. Ercolano for useful discussions concerning disk winds, 
and C. Pinte for discussions about inner disk structure and H$_2$ emission.}

\newpage
\appendix
\section{}

\begin{table*}
\setlength{\tabcolsep}{1.0mm}
\caption{Summary of the observations }              
\centering                                      
\begin{tabular}{c c c c c c c c }          
\hline\hline                        
& & & & & & \multicolumn{2}{c}{PSF FHWM}\\
       $\lambda_{\rm ref}$ & Date   & UT start &  $t_{exp}$  & airmass & slit PA & {\sc science} & {\sc std} \\ 
       $[$nm$]$ &  dd/mm/yyyy & hh:mm     & [s]  & & [$^\circ$] & [mas] & [mas]\\  
\hline      
\multicolumn{8}{c}{{\bf HD~58647} (calibrator: HD~44402)}  \\												
2117.6      	&	27/02/2008	&	03:23	&	160	&	1.1         	&	118  & 220$\pm3$ & 207$\pm5$ \\
2123.3      	&	27/02/2008	&	03:29	&	160	&	1.1         	&	117  & 219$\pm3$ & 218$\pm5$ \\
2236.1      	&	27/02/2008	&	03:34 &	160	&	1.1         	&	116  & 209$\pm3$ & 207$\pm5$ \\
\hline                    												
\multicolumn{8}{c}{{\bf HD~87643} (calibrator: HD~81188)}  \\												
2117.6      	&	14/03/2008	&	04:07 &	32	&	1.2         	&	20  & 232$\pm5$ & 497$\pm24$ \\
2123.3      	&	14/03/2008	&	04:12 &	32	&	1.2         	&	22  & 225$\pm5$ & 387$\pm36$ \\
2236.1      	&	14/03/2008	&	04:17 &	32	&	1.2          	&	24  & 223$\pm8$ & 721$\pm24$ \\
\hline 
\multicolumn{8}{c}{{\bf HD~95881} (calibrator: HIP 061199)}  \\												
2117.7      	&	02/07/2008	&	23:06	&	160	&	1.6         	&	39   & 236$\pm4$ & 256$\pm8$\\
2123.4      	&	02/07/2008	&	23:11	&	160	&	1.6         	&	40   & 214$\pm5$ & 278$\pm9$\\
2236.1     	&	02/07/2008	&	23:53	&	160	&	1.7         	&	52   & 274$\pm45$\tablefootmark{b}  & 237$\pm3$ \\
\hline   
\multicolumn{8}{c}{{\bf HD~97048} (calibrator: HIP 061199)}  \\												
2117.7      	&	04/07/2008	&	23:20	&	160	&	1.8        &	40  & 250$\pm5$ & 256$\pm8$\\
~~2123.3\tablefootmark{a}     &01/04/2007	        &	03:35	&	1920	&	1.7        &	0    & 278$\pm4$ & ...\\
2123.4      	&	04/07/2008	&	23:25	&	160	&	1.8        &	41   & 226$\pm8$ & 278$\pm9$\\
2236.1      	&	05/07/2008	&	00:17	&	240	&	2.0        &	122 & 237$\pm41$ & 237$\pm3$\\
\hline                                           												                    										
\multicolumn{8}{c}{{\bf HD~100546} (calibrator: HIP 061585)}  \\												
2117.7      	&	03/07/2008	&	22:50	&	160	&	1.5         	&	26  & 253$\pm3$ & 209$\pm3$ \\
2123.4      	&	03/07/2008	&	22:55	&	160	&	1.5         	&	28	&	 257$\pm3$ & 202$\pm3$\\
2236.1      	&	03/07/2008	&	23:40	&	160	&	1.5         	&	41 	&	 224$\pm6$ & 228$\pm3$\\
\hline
\multicolumn{8}{c}{{\bf HD~101412} (calibrator: HD~106490)}  \\												
2117.6      	&	15/03/2008	&	06:42	&	240	&	1.3       &	41   & 214$\pm6$ & 540$\pm40$ \\
2123.3      	&	15/03/2008	&	06:47	&	240	&	1.3       &	43   & 206$\pm7$ & 620$\pm31$ \\
2236.1      	&	15/03/2008	&	06:55	&	240	&	1.3       &	46   & 219$\pm30$ & 441$\pm16$ \\
\hline                        												
\multicolumn{8}{c}{{\bf HD~135344B} (calibrator: HD~143118)}  \\												
2117.6      	&	23/02/2008	&	07:44   &	240	&	1.1         	&	-73   & 219$\pm4$ & 213$\pm5$ \\
2123.3      	&	23/02/2008	&	07:50	&	240	&	1.1         	&	-72   & 219$\pm4$ & 203$\pm6$ \\
2236.1      	&	23/02/2008	&	07:56	&	240	&	1.1         	&	-70   & 210$\pm16$ & 200$\pm8$ \\
\hline    
\multicolumn{8}{c}{{\bf HD~141569} (calibrator: HIP 061199)}  \\												
2117.7      	&	03/07/2008	&	02:47	&	160	&	1.1    &	146  & 233$\pm5$ & 256$\pm8$\\
2123.4      	&	03/07/2008	&	02:54	&	160	&	1.1    &	143  & 232$\pm4$ & 278$\pm9$ \\
2236.1      	&	03/07/2008	&	03:33	&	160	&	1.5    &	131  & 213$\pm34$ & 237$\pm3$ \\  
\hline                        												
\multicolumn{8}{c}{{\bf HD~144432} (calibrator: HIP 018266)}  \\												
2117.7      	&	05/07/2008	&	02:35	&	160	&	1.1       &	73  & 214$\pm3$ & 220$\pm3$ \\
2123.4      	&	05/07/2008	&	02:40	&	160	&	1.1       	&	76  & 218$\pm4$ & 230$\pm3$ \\
2236.1      	&	05/07/2008	&	03:18	&	160	&	1.1       	&	86  & 203$\pm25$ & 229$\pm3$ \\
\hline                        												
\multicolumn{8}{c}{{\bf HD~150193} (calibrator: HIP 090422)}  \\												
2117.7      	&	05/07/2008	&	05:01	&	160	&	1.2        &	99   & 216$\pm5$ & 192$\pm4$ \\
2123.4      	&	05/07/2008	&	05:07	&	160	&	1.2        &	100  & 217$\pm3$ & 197$\pm4$ \\
2236.1      	&	05/07/2008	&	05:50	&	160	&	1.4        &	102  & 231$\pm27$ & 206$\pm4$ \\
\hline                          												
\multicolumn{8}{c}{{\bf 51 Oph} (calibrator: HIP 081266)}  \\												
2117.7      	&	04/07/2008	&	02:25	&	80	&	1.0        	&	-96  & 220$\pm3$ & 221$\pm3$	\\
2123.4      	&	04/07/2008	&	02:30	&	80	&	1.0         	&	-96  & 220$\pm3$ & 230$\pm3$ \\
2236.1      	&	04/07/2008	&	03:06  &	80	&	1.0         	&	-101 & 215$\pm5$ & 229$\pm3$	\\
\hline    
\multicolumn{8}{c}{{\bf HD~169142} (calibrator: HIP 095347)}  \\												
2117.7      	&	04/07/2008	&	04:58	&	160	&	1.0         	&	63  & 208$\pm3$ & 230$\pm3$ \\
2123.4      	&	04/07/2008	&	05:03	&	160	&	1.0         	&	66  & 219$\pm4$ & 208$\pm3$ \\
2236.1      	&	04/07/2008	&	05:45	&	160	&	1.1         	&	81  & 216$\pm27$ & 217$\pm6$ \\
\hline	
\multicolumn{8}{c}{{\bf R CrA} (calibrator: HIP 095347)}  \\												
2117.7      	&	04/07/2008	&	08:15	&	40	&	1.4         	&	89  & 219$\pm5$ & 230$\pm3$ \\
2123.4      	&	04/07/2008	&	08:19	&	40	&	1.4         	&	89  & 218$\pm3$ & 208$\pm3$ \\
2236.1      	&	04/07/2008	&	08:54	&	40	&	1.6         	&	94  & 219$\pm25$ & 217$\pm6$ \\
\hline                        												                        												
\multicolumn{8}{c}{{\bf HD~179218} (calibrator: HIP 112029)}  \\												
2117.7      	&	03/07/2008	&	05:29	&	160	&	1.3          	&	172  & 218$\pm3$ & 225$\pm6$ \\
2123.4      	&	03/07/2008	&	05:34	&	160	&	1.3         &	170  & 222$\pm2$ & 217$\pm3$ \\
2236.1      	&	03/07/2008	&	06:17	&	160	&	1.4         &	157  & 222$\pm33$ & 220$\pm2$ \\                      \hline                												                      												
\multicolumn{8}{c}{{\bf HD~190073} (calibrator: HIP 112029 \& HIP 090422)}  \\									
2117.7      	&	03/07/2008	&	08:22	&	160	&	1.5         	&	132  & 220$\pm6$ & 192$\pm4$ \\
2123.4      	&	03/07/2008	&	08:27	&	160	&	1.5         	&	131  & 220$\pm3$ & 197$\pm3$ \\
2236.1      	&	05/07/2008	&	09:29	&	160	&	1.8         	&	125  & 223$\pm35$ & 206$\pm4$ \\
\hline    
\end{tabular}
\flushleft
\tablefoot{
\tablefoottext{a}  Observations from ESO program 079.C-0860(C): (i) observations with the slit aligned north-south; 
(ii) No suitable PSF standard observed at the night of the observations,
we used a standard star from our program for the flux calibration.
\tablefoottext{b} Bad weather conditions during the observation.}

\label{table_observations}
\end{table*}

\begin{figure*}
   \centering                  
\begin{tabular}{cc}
\includegraphics[width=0.5\textwidth]{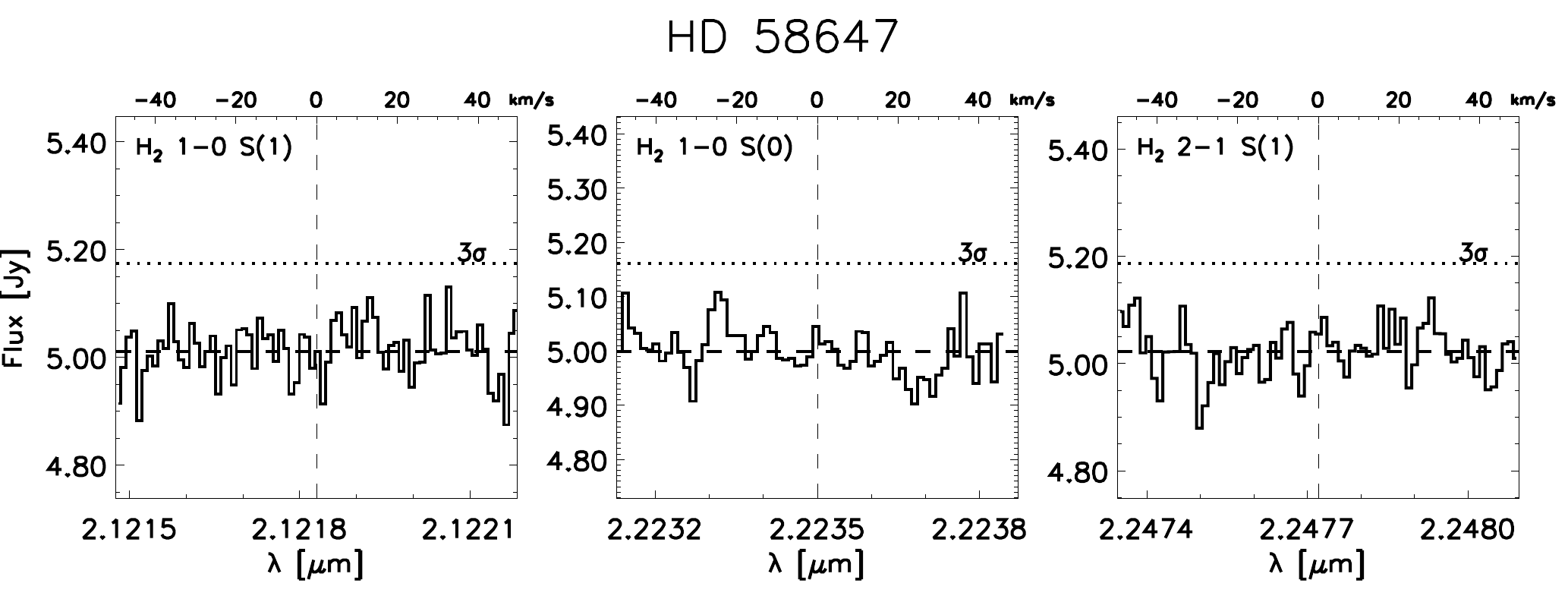}&
\includegraphics[width=0.5\textwidth]{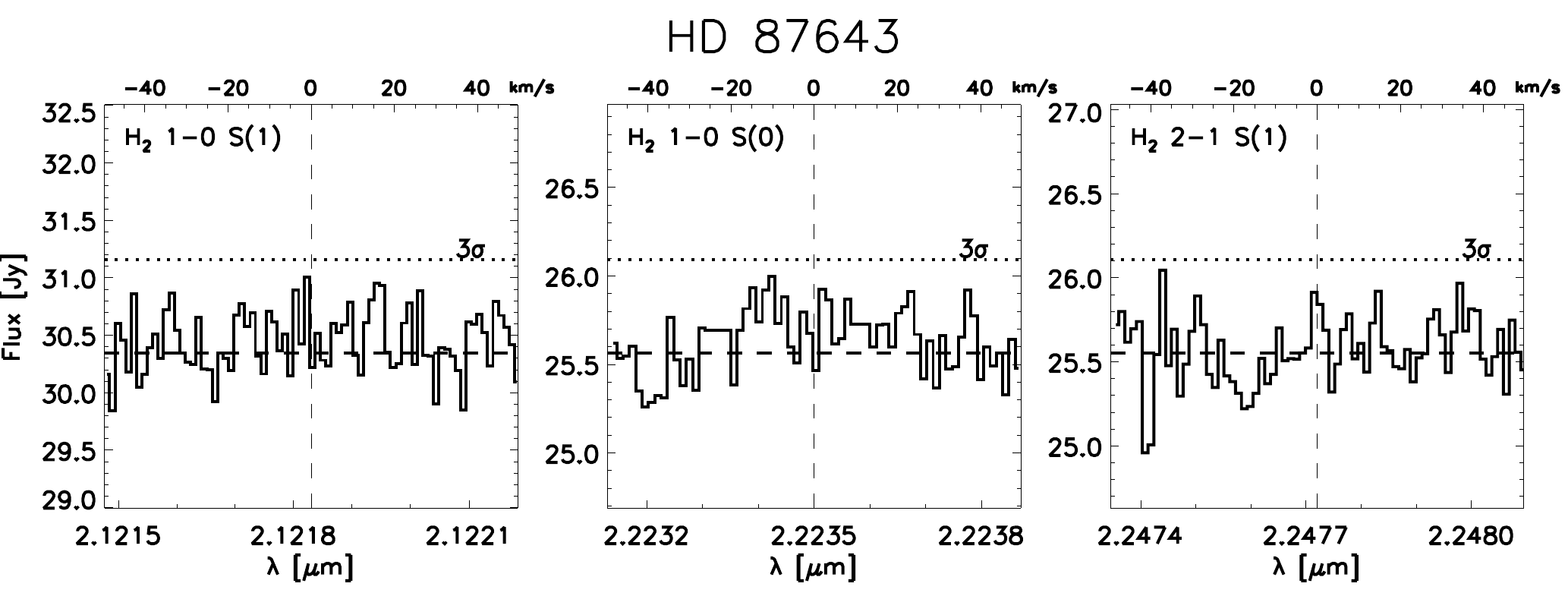} \\
\includegraphics[width=0.5\textwidth]{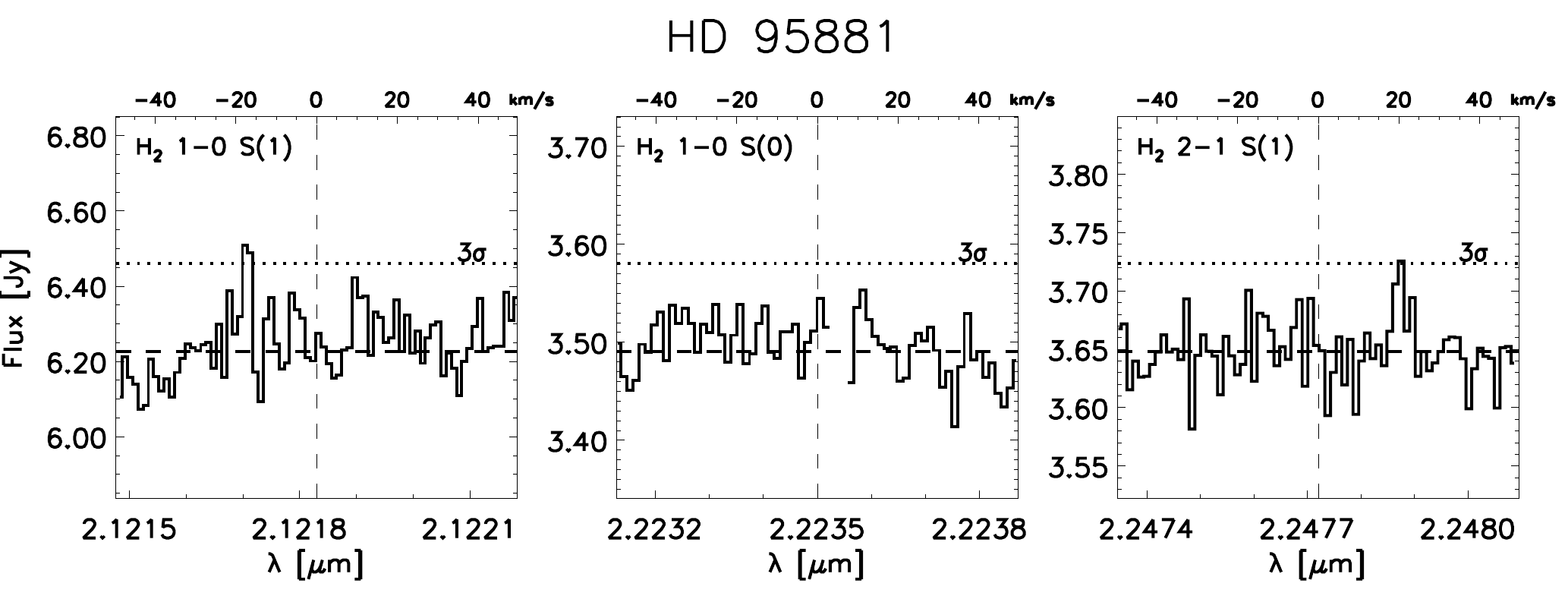} &
\includegraphics[width=0.5\textwidth]{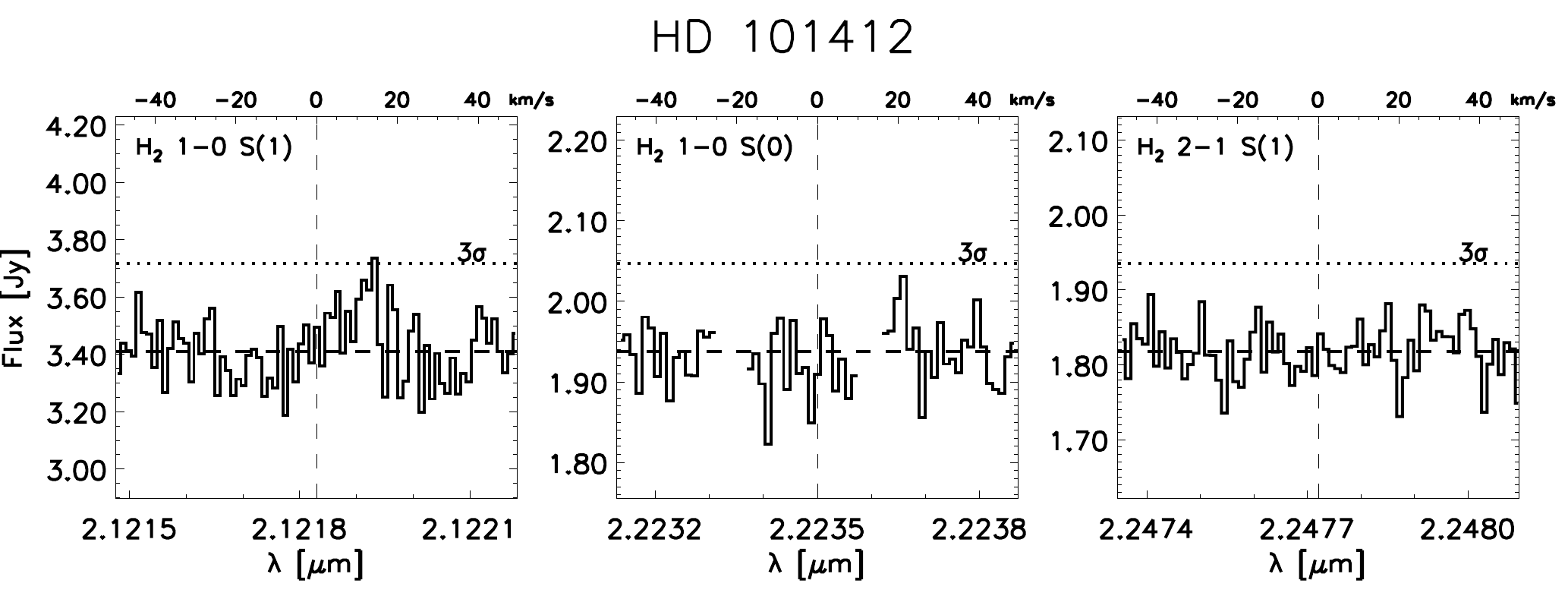} \\
\includegraphics[width=0.5\textwidth]{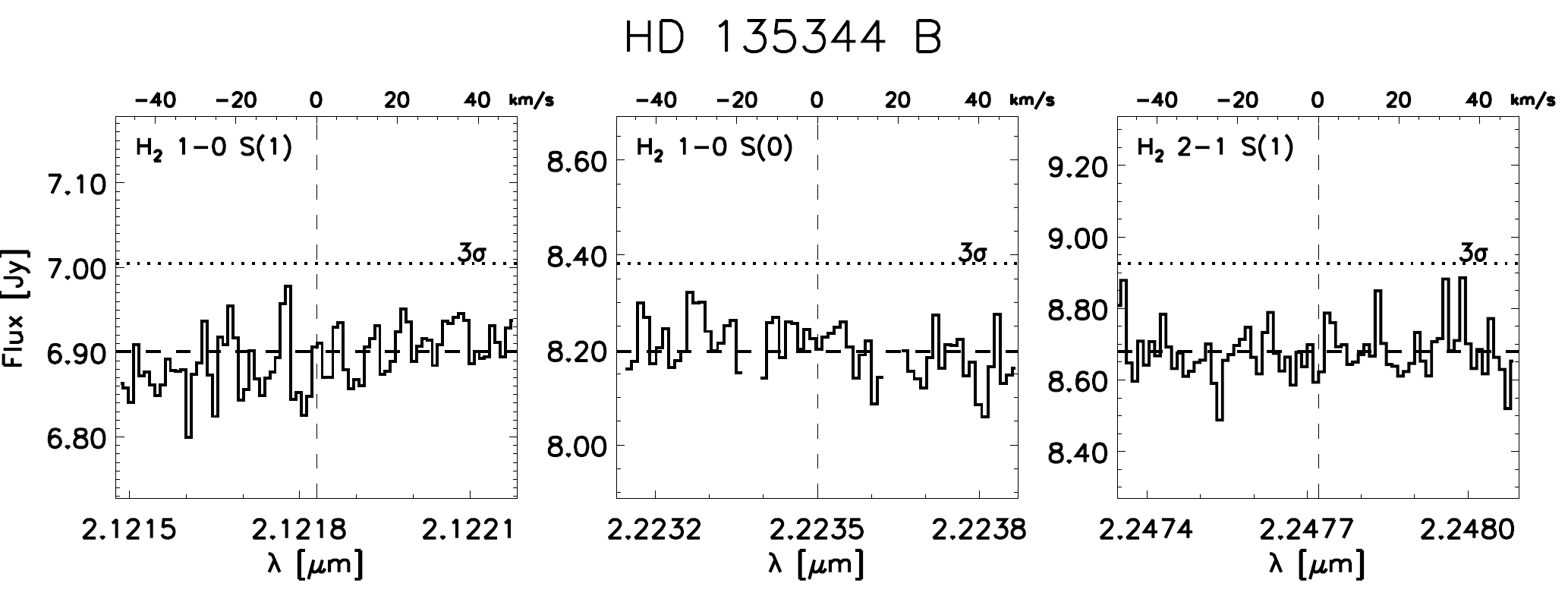} &
\includegraphics[width=0.5\textwidth]{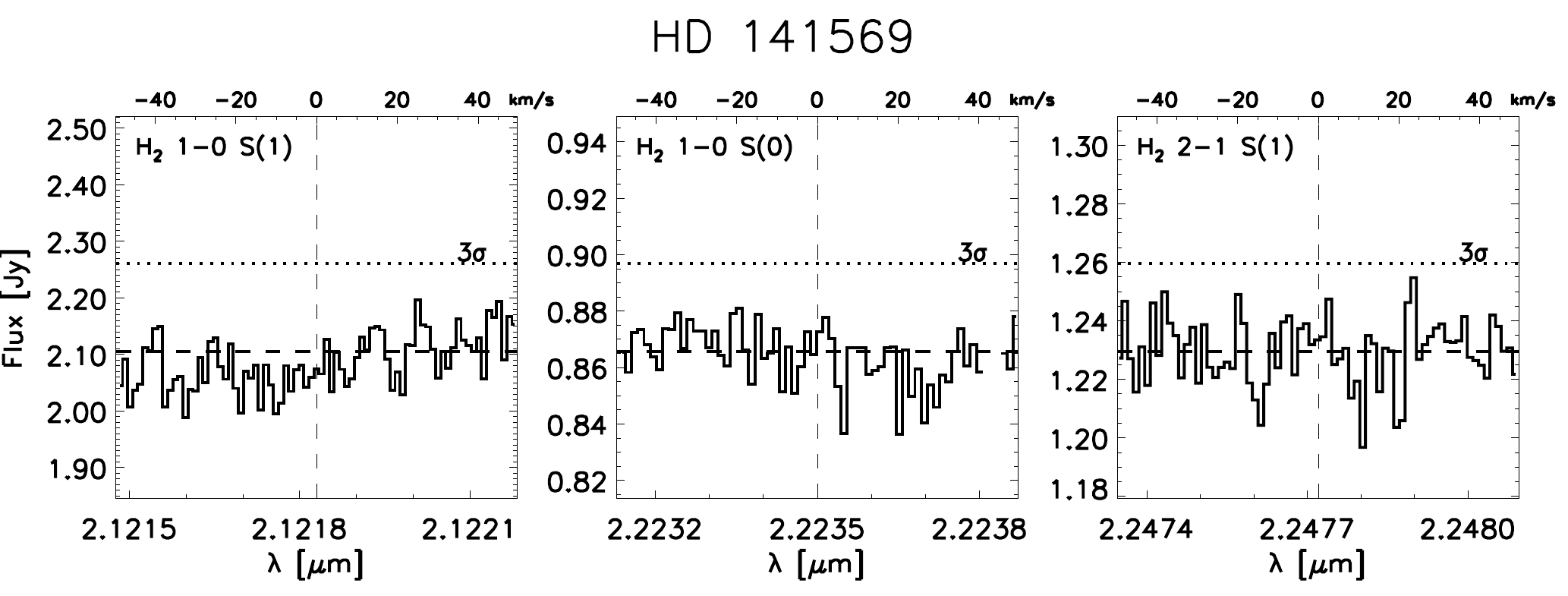} \\
\includegraphics[width=0.5\textwidth]{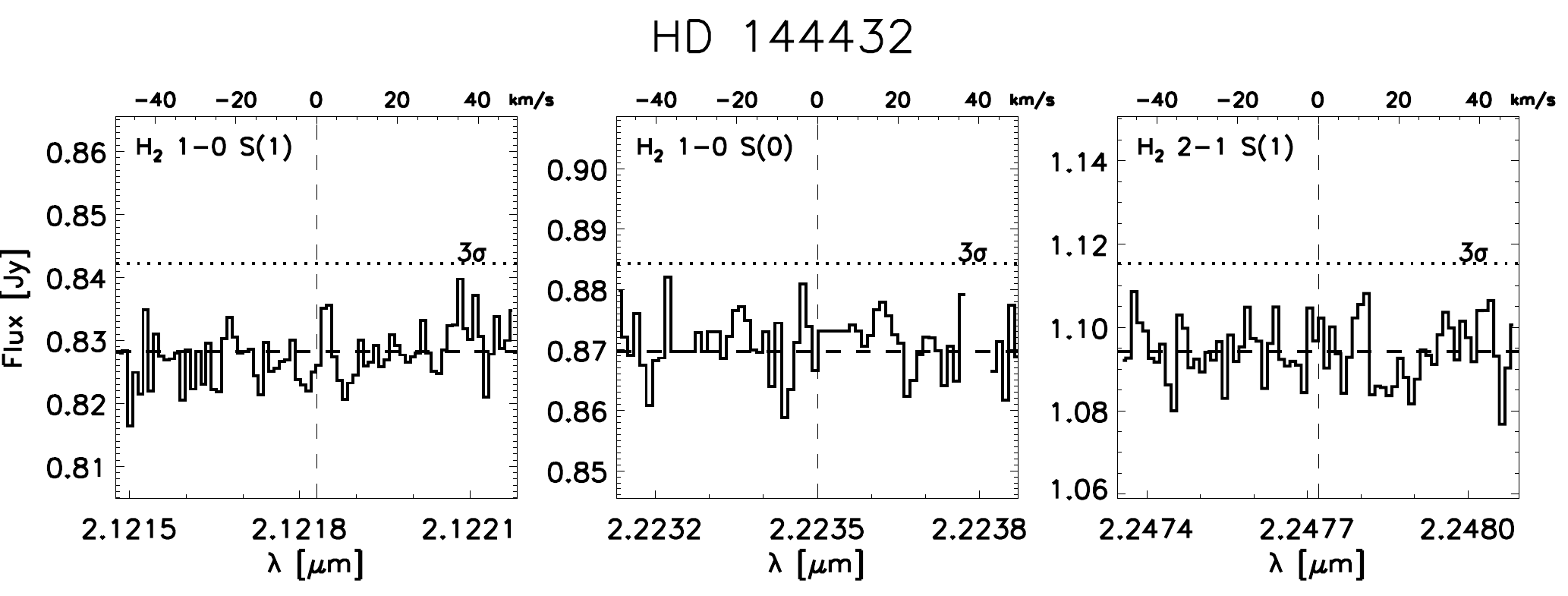} &
\includegraphics[width=0.5\textwidth]{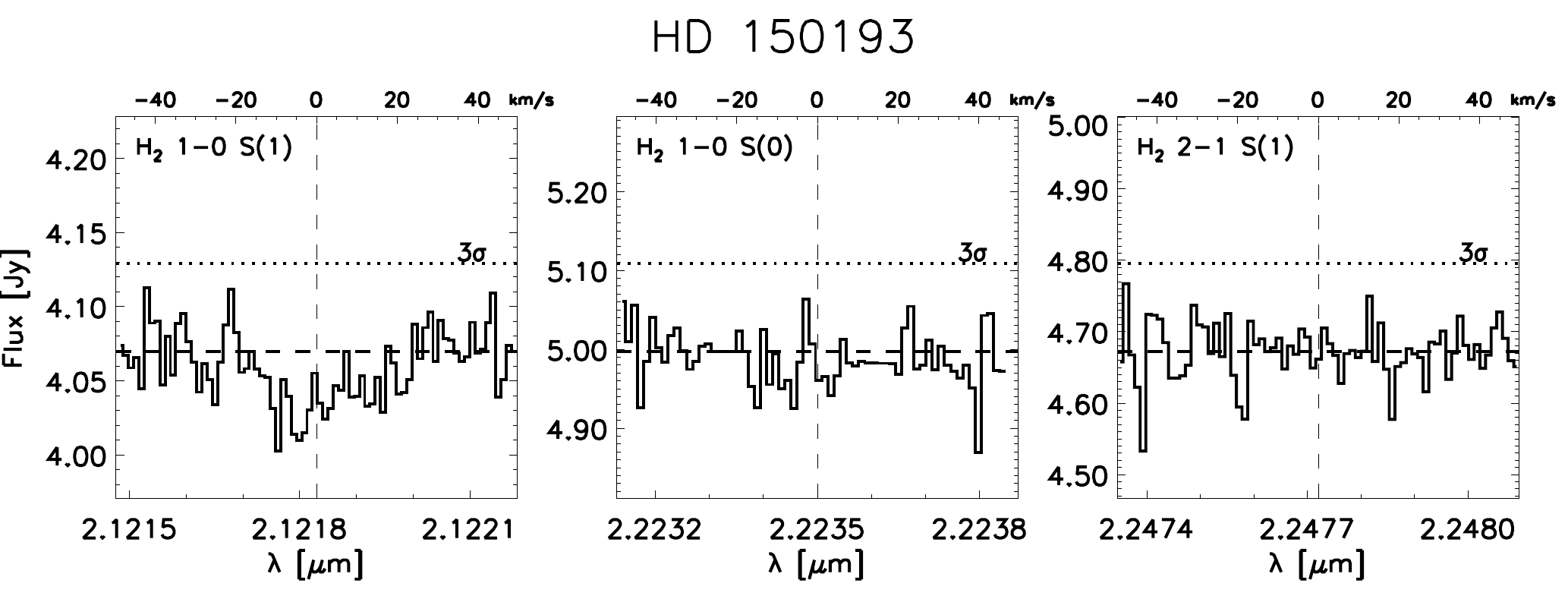} \\
\includegraphics[width=0.5\textwidth]{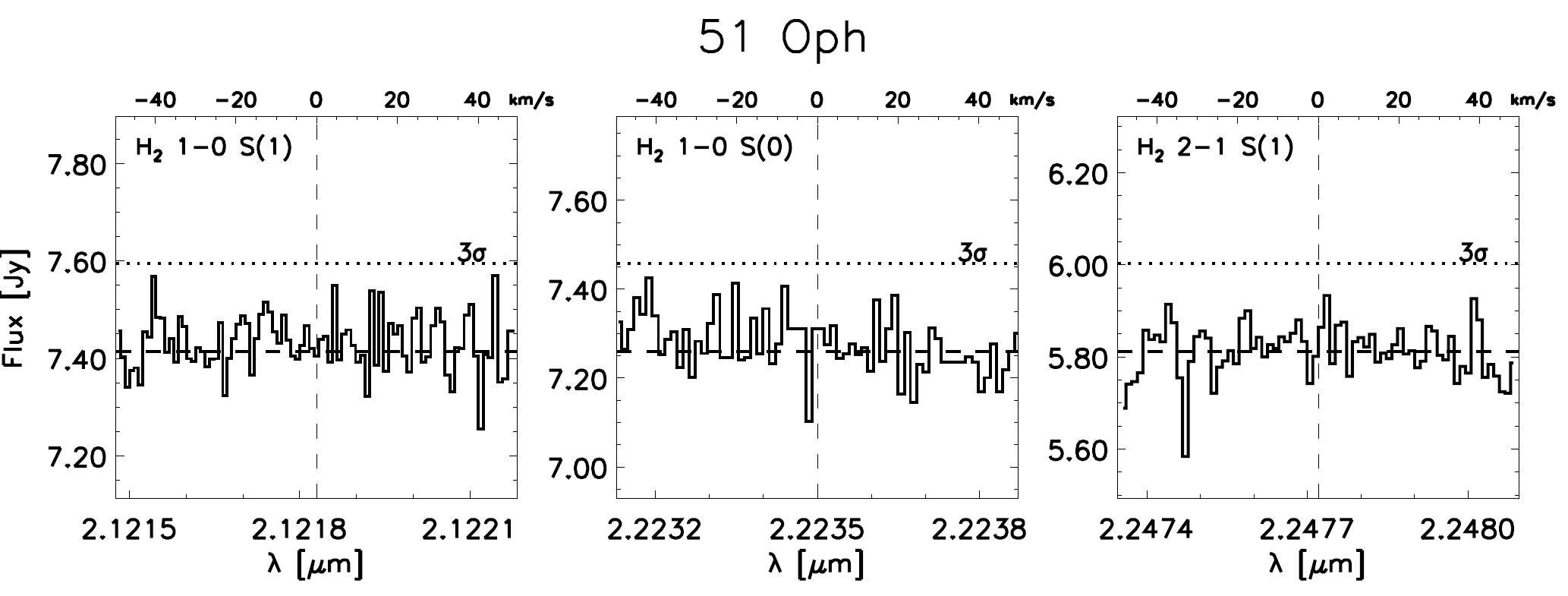} &
\includegraphics[width=0.5\textwidth]{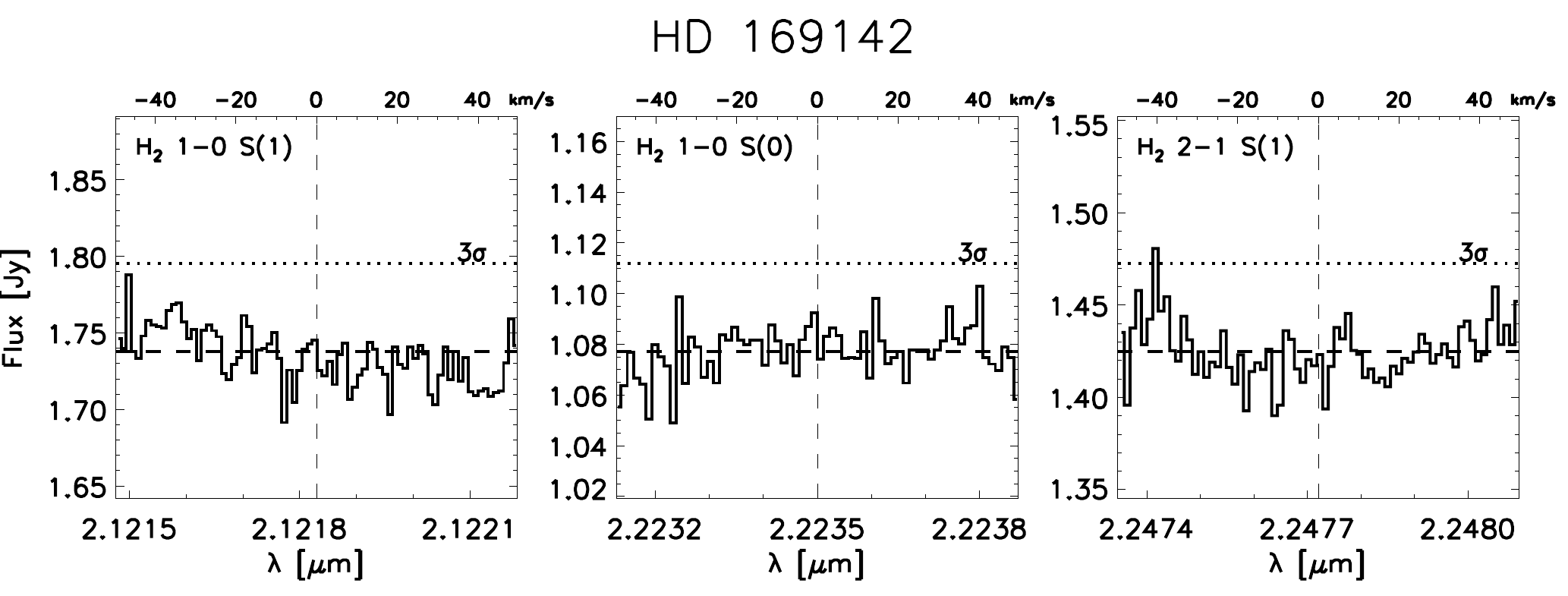} \\
\includegraphics[width=0.5\textwidth]{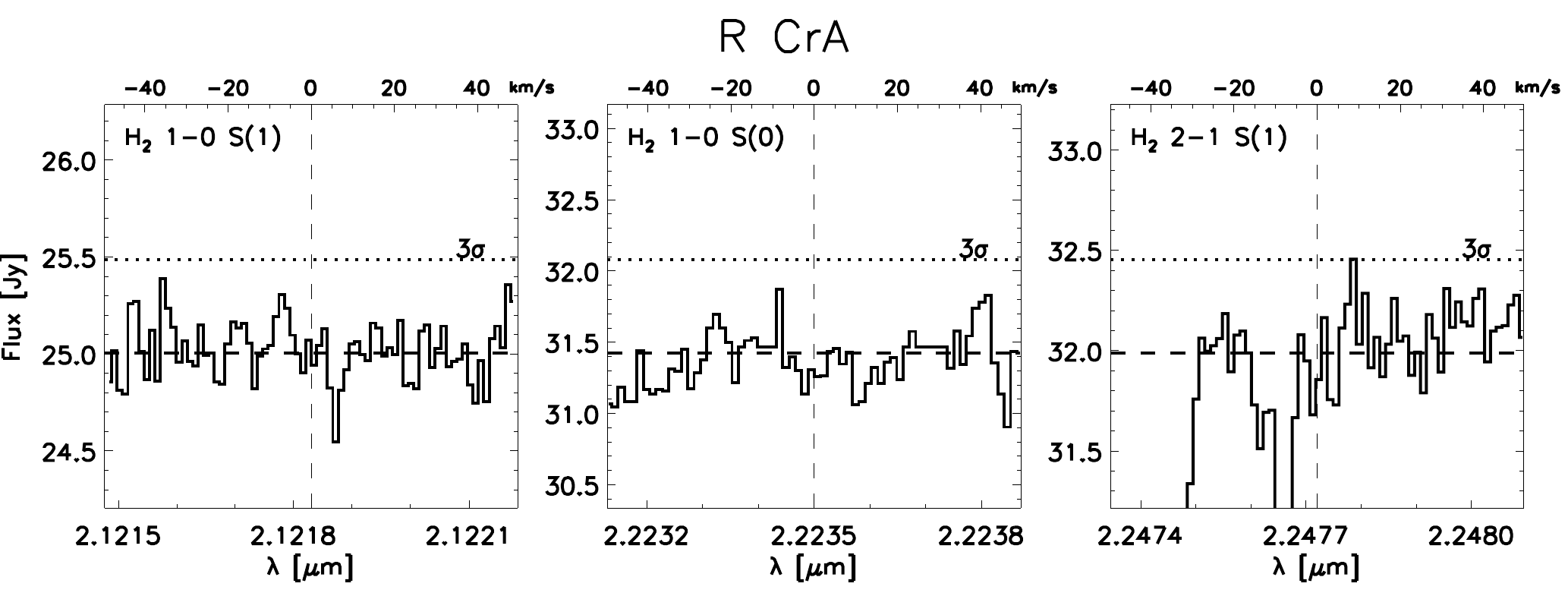} &
\includegraphics[width=0.5\textwidth]{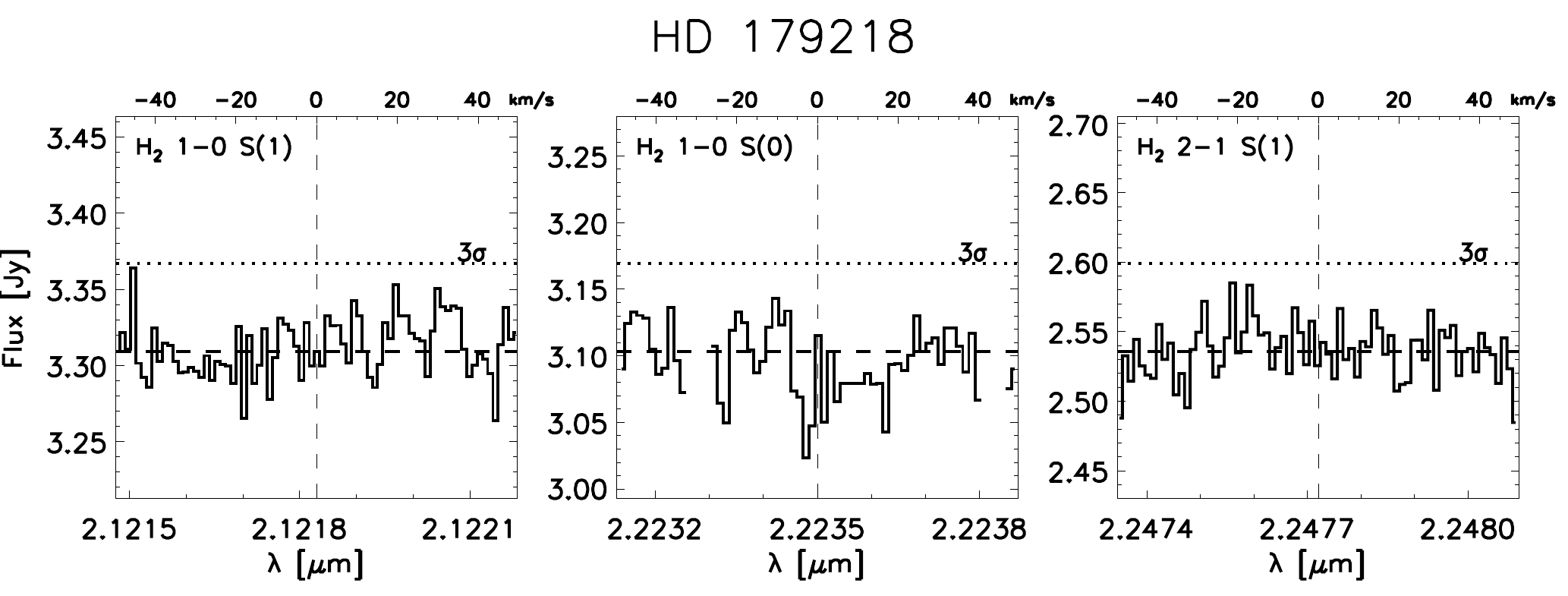}\\
\includegraphics[width=0.5\textwidth]{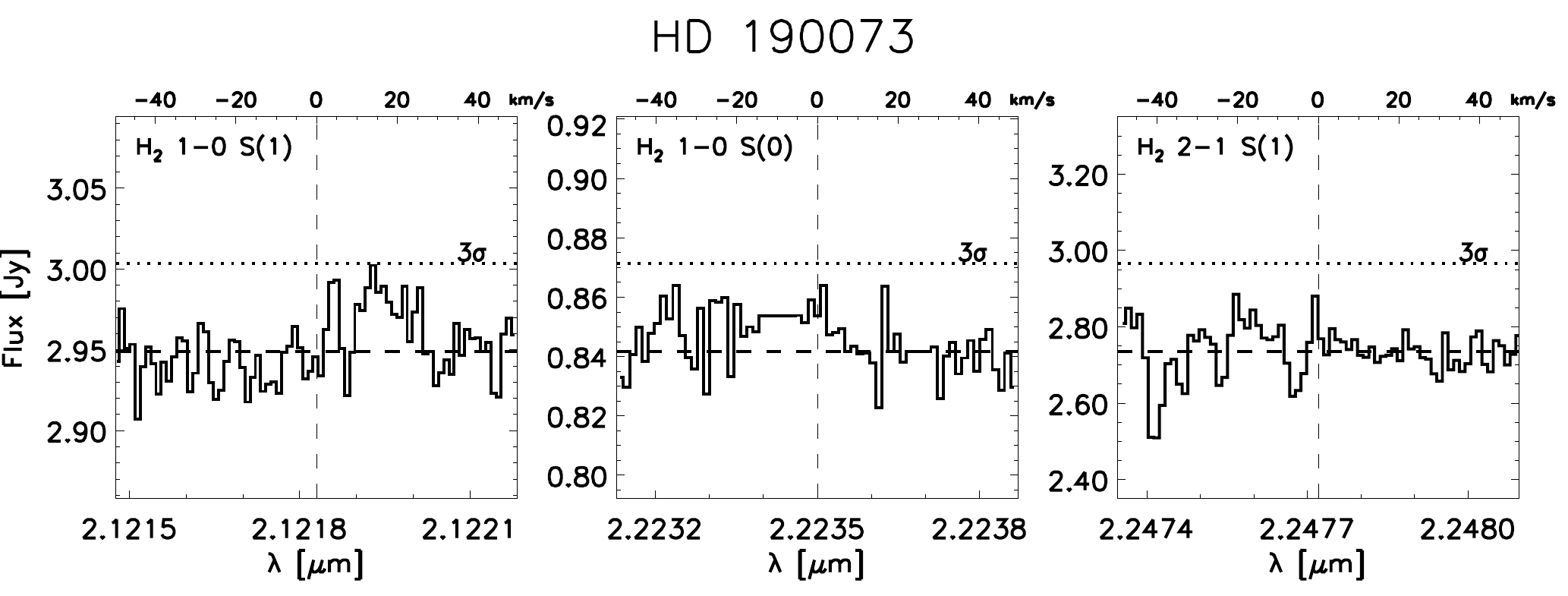}
\end{tabular}
\caption{Observed CRIRES spectrum at the location of the H$_2$ 1-0 S(1), 1-0 S(0) and 2-1 S(1) lines in the
13 targets in which H$_2$ emission is not detected. 
The spectra are presented at the rest velocity of the stars.
Upper limits to the line fluxes are given in Table \ref{upper_limits}. 
}
\label{fig_nondetections}
\end{figure*} 

\end{document}